%% file: main.tex
\documentclass[preprint2]{aastex6}

\newcommand{\Ka}{Fe\,{\small I}-K$\alpha$}
\newcommand{\Kb}{Fe\,{\small I}-K$\beta$}
\newcommand{\Hea}{Fe\,{\small XXV}-He$\alpha$}
\newcommand{\Heb}{Fe\,{\small XXV}-He$\beta$}
\newcommand{\Hec}{Fe\,{\small XXV}-He$\gamma$}
\newcommand{\Lya}{Fe\,{\small XXVI}-Ly$\alpha$}
\newcommand{\Lyb}{Fe\,{\small XXVI}-Ly$\beta$}
\newcommand{\Lyc}{Fe\,{\small XXVI}-Ly$\gamma$}
\newcommand{\NiKa}{Ni\,{\small I}-K$\alpha$}
\newcommand{\NiHea}{Ni\,{\small XXVII}-He$\alpha$}
\newcommand{\EWKa}{EW$_{6.40}$}
\newcommand{\SHKa}{SH$_{6.40}$}
\newcommand{\EWHea}{EW$_{6.68}$}
\newcommand{\SHHea}{SH$_{6.68}$}
\newcommand{\EWLya}{EW$_{6.97}$}
\newcommand{\SHLya}{SH$_{6.97}$}
\newcommand{\Fu}{\,erg\,cm$^{-2}$\,s$^{-1}$}
\newcommand{\Lu}{\,erg\,s$^{-1}$}
\newcommand{\NH}{$N_{\rm H}$}

\newcommand{\KaTab}{Fe{\scriptsize I}-K$\alpha$}
\newcommand{\KbTab}{Fe{\scriptsize I}-K$\beta$}
\newcommand{\HeaTab}{Fe{\scriptsize XXV}-He$\alpha$}
\newcommand{\HebTab}{Fe{\scriptsize XXV}-He$\beta$}
\newcommand{\HecTab}{Fe{\scriptsize XXV}-He$\gamma$}
\newcommand{\LyaTab}{Fe{\scriptsize XXVI}-Ly$\alpha$}
\newcommand{\LybTab}{Fe{\scriptsize XXVI}-Ly$\beta$}
\newcommand{\LycTab}{Fe{\scriptsize XXVI}-Ly$\gamma$}
\newcommand{\NiKaTab}{Ni{\scriptsize I}-K$\alpha$}
\newcommand{\NiHeaTab}{Ni{\scriptsize XXVII}-He$\alpha$}

\begin{document}

\title{Origin of the Galactic Diffuse X-ray Emission: Iron K-shell Line Diagnostics}

\author{
	Masayoshi Nobukawa\altaffilmark{1}, 
	Hideki Uchiyama\altaffilmark{2},
	Kumiko K. Nobukawa\altaffilmark{3,4},
	Shigeo Yamauchi\altaffilmark{4}, and
	Katsuji Koyama\altaffilmark{3,5}}

\altaffiltext{1}{Department of Teacher Training and School Education, Nara University of Education, Takabatake-cho, Nara,  630-8528, Japan}
\altaffiltext{}{nobukawa@nara-edu.ac.jp}
\altaffiltext{2}{Faculty of Education, Shizuoka University, 836 Ohya, Suruga-ku, Shizuoka, 422-8529, Japan}
\altaffiltext{3}{Department of Physics, Graduate School of Science, Kyoto University, Kitashirakawa-oiwake-cho, Sakyo-ku, Kyoto, 606-8502, Japan}
\altaffiltext{4}{Department of Physics, Nara Women's University, Kitauoyanishimachi, Nara, 630-8506, Japan}
\altaffiltext{5}{Department of Earth and Space Science, Graduate School of Science, Osaka University, 1-1 Machikaneyama-cho, Toyonaka, Osaka, 560-0043, Japan}

\begin{abstract}

This paper reports detailed K-shell line profiles of iron (Fe) and nickel (Ni) of the Galactic Center X-ray Emission (GCXE), Galactic Bulge X-ray Emission (GBXE), Galactic Ridge X-ray Emission (GRXE), magnetic Cataclysmic Variables (mCVs), non-magnetic Cataclysmic Variables (non-mCVs), and coronally Active Binaries (ABs). For the study of the origin of the GCXE, GBXE, and GRXE, the spectral analysis is focused on equivalent widths of the \Ka, \Hea, and \Lya~lines. The global spectrum  of the GBXE is reproduced by a combination of the mCVs, non-mCVs, and ABs spectra. On the other hand,  the GRXE spectrum shows significant data excesses at the \Ka\ and \Hea~line energies. This means that additional components other than mCVs, non-mCVs, and ABs are required, which have symbiotic phenomena of cold gas and very high-temperature plasma. The GCXE spectrum shows larger excesses than those found in the GRXE spectrum at all the K-shell lines of iron and nickel. Among them the largest ones are the \Ka, \Hea, \Lya, and \Lyb~lines. Together with the fact that the scale heights of the \Ka, \Hea, and \Lya\ lines are similar to that of the central molecular zone (CMZ), the excess components would be related to high-energy activity in the extreme envelopment of the CMZ. 

\end{abstract}


\section{Introduction} \label{sec:intro}
The Galactic Diffuse X-ray Emission (GDXE) is unresolved X-rays prevailing over the Galactic plane (e.g. \citealt{Wo82}). One of the most remarkable features of the 
GDXE is strong K-shell lines of neutral (\Ka), helium-like (\Hea), and hydrogen-like irons (\Lya) at 6.40 keV, 6.68 keV, and 6.97 keV, respectively \citep{Ko96,Ko07d,Ya09}.  
The GDXE is decomposed into the Galactic Center X-ray Emission (GCXE), the Galactic Bulge X-ray Emission (GBXE), and the Galactic Ridge X-ray Emission (GRXE) 
\citep{Ko89, Ya93, Uc13, Ya16}. Although the global X-ray spectra of the GCXE, GBXE, and GRXE are similar to each other, the detailed structures, particularly the iron K-shell line structures are significantly different \citep{Ya16}. 

Since the discovery of the GDXE, its origin, whether unresolved  point sources or truly diffuse plasma, has been under debate for a long time. In the point source scenario, the
candidate stars have been mainly Cataclysmic Variables (CVs) and coronally Active Binaries (ABs) with their reasonable number densities, high-temperature 
plasma, and strong iron K-shell lines \citep{Re09, Wa14b}. 
The previous debates around the point source origin have been based on two observational facts that (1) the longitude distributions of the continuum 
(e.g. 2--10 keV band) and the iron K-shell line fluxes resemble to the infrared surface brightness distribution \citep{Re06a,Re06b}, which is regarded as a tracer of  the stellar mass distribution (SMD), and (2) the flux of the resolved point sources is roughly equal to the total GDXE flux, 
if the reliable point source flux in the luminosity range of $\gtrsim10^{31}$~\Lu~is extrapolated to the low-luminosity limit of $\sim10^{28}$~\Lu~using empirically made X-ray luminosity functions (XLFs) \citep{Sa06, Re09, Wa14b}.

These processes, however, have large observational uncertainty. 
In (1), both the X-ray and infrared SMDs are made with poor spatial sub-degree resolution. Therefore, the fine profiles of the GCXE 
of $\sim 1\degr.2 \times 0\degr.5$ size, and the boundaries between the GCXE, GBXE, and GRXE are smeared out.  The latitude distribution of the GDXE flux is not 
also determined precisely. The infrared surface brightness distribution is a tracer of the star distribution including high-mass objects, but it is not clear how correctly it traces the distribution of low-mass stars such as CVs and ABs. 
In (2), the XLFs are made by limited sample numbers, luminosity range, and spectral information of candidate  point sources, and hence
have  large  errors of $\gtrsim$50\%  (e.g. \citealt{Sa06}). In fact, the XLFs are largely different from author to author \citep{Re09, Ho12, Wa14b}.
In addition, the resolved fraction, even in the reliable luminosity range of $\gtrsim10^{31}$~\Lu, is typically $\sim$10--30\% with uncertainty of  factor of $\sim3$. 

In the examination of the GDXE origin, whether from point sources, truly diffuse plasma or other origins, equivalent widths (EWs) of  
\Ka~(\EWKa), \Hea~(\EWHea), and \Lya~(\EWLya) of the GDXE, magnetic CVs (mCVs), non-magnetic CVs (non-mCVs), and ABs are key factors.
The limited energy resolutions of previous observations cannot separate the iron K-shell lines into  \Ka, \Hea\, and \Lya, and hence the EWs of the GCXE, GBXE, and GRXE have not been accurate enough. In addition, the EWs of mCVs, non-mCVs and ABs have also been very limited, with large errors or significant variations from author to author, and/or from instrument to instrument. 
The best-quality global spatial and spectral structures of the GCXE, GBXE, and GRXE, and high-quality spectra of mCVs, non-mCVs, and ABs become available from the {\it Suzaku} observations: see \citet{Ko07b, Uc11} for the GDXE, and \citet{Xu16} for mCVs, non-mCVs and ABs.

The motivation of this work is to try different diagnostics from the previous methods for the origins of the GCXE, GBXE, and GRXE.  Our new diagnostics is to use high-quality spectra with accurate \EWKa, \EWHea, and \EWLya\ from the GCXE, GBXE, GRXE, mCVs, non-mCVs, and ABs, which are obtained from all the {\it Suzaku} archives. This approach is applied partly by \citet{Xu16}, and 
this paper intends to further extend their method.

The contents of this paper is as follows. The observations and data reductions are described in section~\ref{sec:obs}.
The EWs of the K-shell lines of iron and nickel from mCVs, non-mCVs, ABs, and the GDXE obtained from all the available {\it Suzaku} archives are presented in sections~\ref{sec:GDXEspectra} and \ref{sec:AS}. Mean spectra of mCVs, non-mCVs, and ABs are constructed in sections~\ref{sec:modelA} and \ref{sec:modelB}.
In section~\ref{sec:reconstruction}, we fit the GDXE spectra with a combination of the mean spectra, focusing on the K-shell line structures.  
Using the results, the origins of the GCXE, GRXE, and GBXE are separately discussed in section~\ref{sec:discussion}.

\section{Observations and Data Reduction} \label{sec:obs}
The {\it Suzaku} archives are the data taken in the full mission life of {\it Suzaku} from 2005 to 2015. We used the X-ray Imaging Spectrometers (XIS, \citealt{Ko07a}) placed on the focal  planes of the thin-foil X-ray telescopes (\citealt{Se07}).
The XIS consists  of four sensors: XIS sensor-1 (XIS\,1) has a back-illuminated CCD, 
while the other three XIS sensors (XIS\,0, 2, and 3) have  front-illuminated CCDs.
XIS\,2 turned dysfunctional, and hence the other three sensors (XIS\,0, 1, and 3) have been operated since  2006 November 9.
A small fraction of the XIS\,0 area has not been used since 2009 June 23 because of the  damage by a possible micro-meteorite.
The XIS has been operated in the normal clocking mode. The field of view of the XIS is $17'.8 \times 17'.8$.

Data reduction and analysis were carried out using the HEAsoft version 6.17.
The XIS pulse-height data for each X-ray event were converted to pulse invariant  channels using the {\tt xispi} software and the calibration database.
The data obtained at the South Atlantic Anomaly, during Earth occultation, and at low elevation angles 
from the Earth rim of $<$ 5$^{\circ}$ (night Earth) or $<20^{\circ}$ (day Earth) were excluded. 
After removing hot and flickering pixels, the events  of  grade 0, 2, 3, 4, and 6 were used. 

\begin{figure*}[hbt]
\figurenum{1}
\epsscale{.5}
\plotone{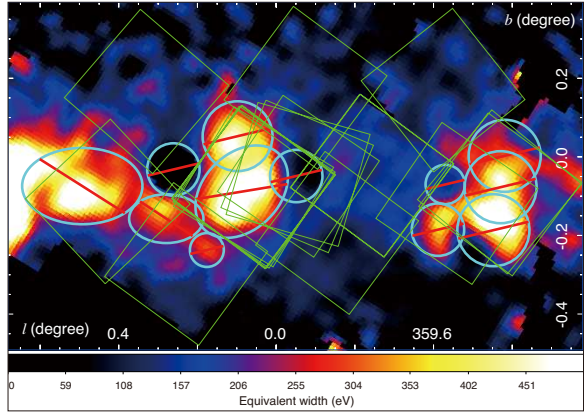}
\caption{The regions of the GCXE spectrum overlaid on the \Ka~EW map. 
The spectrum of the GCXE is extracted from the green squares excluding
the light blue circles.}
\label{fig:GCXE-region}
\end{figure*}

\section{Analysis and Results} 
\label{sec:ana}
\subsection {X-ray spectra of the GCXE, GBXE, and GRXE} 
\label{sec:GDXEspectra}

\input{table-GCXE-Uc.tex} 

\begin{figure*}[hptb]
\figurenum{2}
\epsscale{0.33}
\plotone{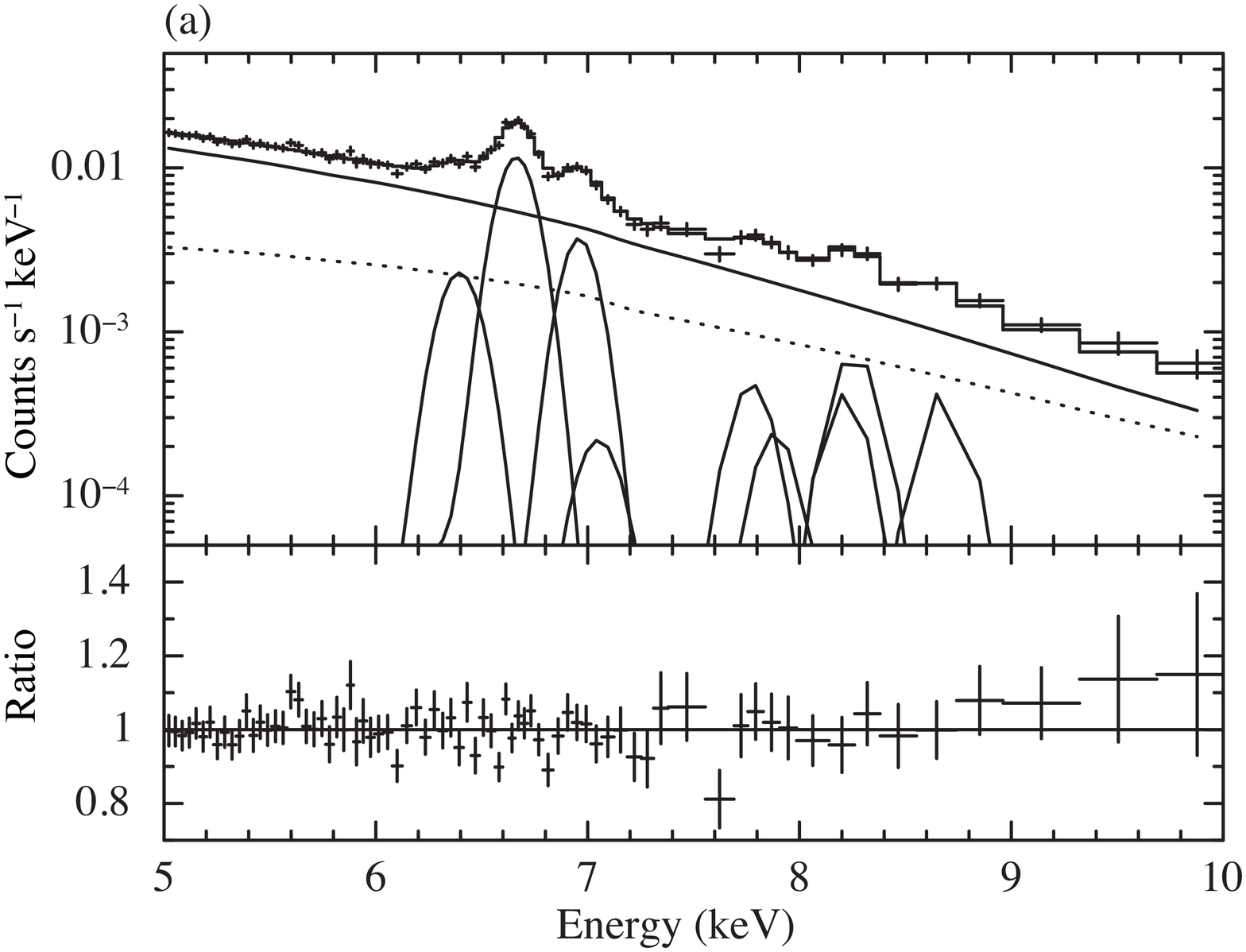}
\plotone{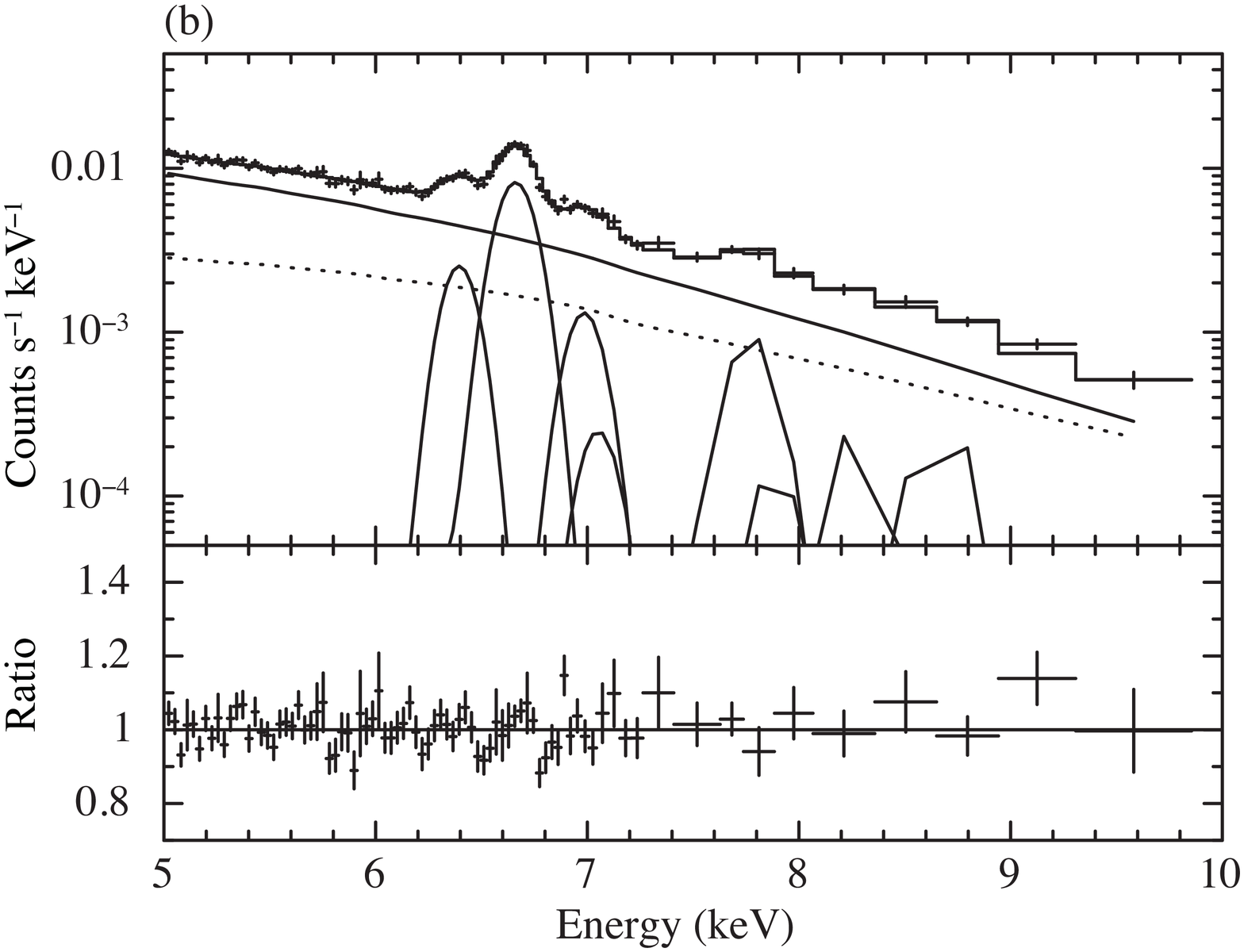}
\plotone{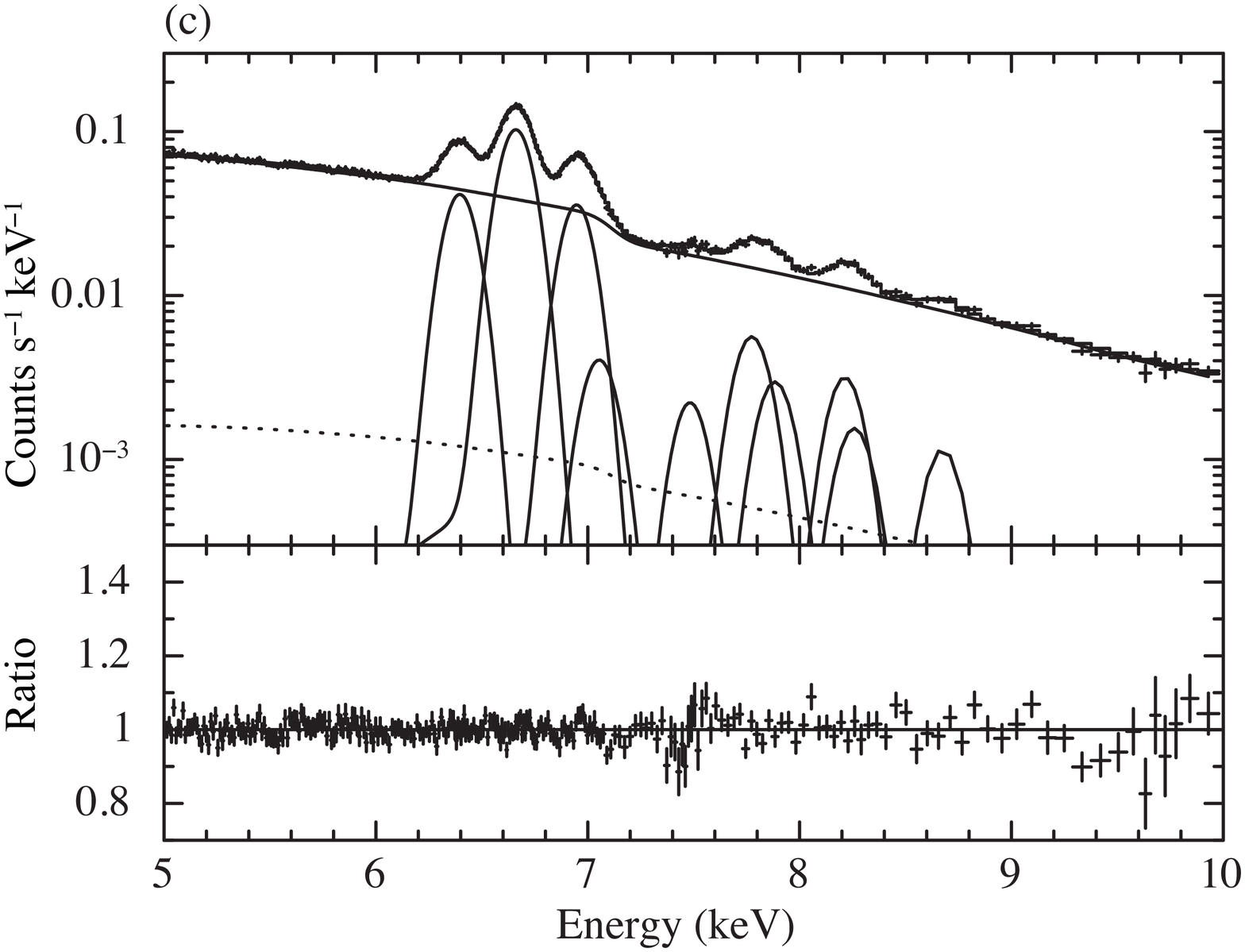}
\caption{
GDXE spectra for the GBXE (a), GRXE (b), and GCXE (c). The spectra are fitted with a phenomenological model consisting of 
bremsstrahlung and ten Gaussians with intersteller absorption. 
The CXB is added with the dotted model.
The best-fit parameters are listed in table~\ref{tab:GDXE}.
}
\label{fig:Spectra}
\end{figure*}

Following the results of \citet{Uc13} and \citet{Ya16}, we selected the data of the GCXE, GBXE, and GRXE from the  ($|l_{*}|< 0\degr.6 $, $|b_{*}|< 0\degr.25$), ($|l_{*}|<0\degr.6$, $1\degr<|b_{*}|< 3\degr$) and ($|l_{*}|=10\degr$--$30\degr$, $|b_{*}|<1\degr$) regions, respectively.  Here we define a new coordinate of $l_*$ = $l+ 0\degr.056$ and  
$b_*$=$b+ 0\degr.046$, where the position of Sagittarius (Sgr) A$^*$ is given by  $(l_*, b_*)=(0\degr, 0\degr)$ \citep{Re04}.  
To obtain the pure GCXE spectrum, bright \Ka\ and \Hea~spots of Sgr A, B, C, Sgr A East, and the Arches Cluster
\citep{Ko07b, Ko07c, Ts07, Na09, No10}, and bright LMXBs 
(1E~1743.1$-$2843 and AX~J1744.8$-$2921, \citealt{Va81,Sa02}) were excluded. 
The region of the GCXE is shown in figure~\ref{fig:GCXE-region}. The total exposure times are $\sim$1.3~Ms, $\sim$800~ks, and $\sim$3.0~Ms, for the GCXE, GBXE, and GRXE, respectively. 

We then made raw spectra from all the GCXE, GBXE, and GRXE regions. The non-X-ray background (NXB) was subtracted using the {\tt xisnxbgen} software \citep{Ta08}. 
We fit the NXB-subtracted spectra with a phenomenological model consisting of a bremsstrahlung continuum, an iron K-shell absorption edge ({\tt edge} in XSPEC), and many Gaussian lines plus the cosmic X-ray background (CXB).
 The Gaussian line energies are taken from the 
AtomDB~3.0.2\footnote{http://www.atomdb.org/}, 
while the widths of the \Hea~lines and the others are  fixed to $\sim30$~eV and $\sim$0 eV, respectively.

The CXB is given  by a power-law function with fixed photon index and flux of 1.4 and  $8.2\times10^{-7}$ photons s$^{-1}$~cm$^{-2}$~arcmin$^{-2}$ keV$^{-1}$ at 1~keV, respectively \citep{Ku02}.
The best-fit results are shown in figure~\ref{fig:Spectra}, while the best-fit parameters are listed in table~\ref{tab:GDXE}. 

The ratio of \Lya/\Hea~of the GCXE, GBXE, and GRXE are $0.37\pm{0.02}$,  $0.34\pm{0.03}$, and $0.17\pm{0.02}$, which correspond to the collisional ionization equilibrium (CIE) temperatures of $\sim6.8$, $\sim6.5$, and $\sim5.0$\,keV, respectively. Thus, the temperatures of the GCXE, GBXE, and GRXE are not significantly different from each other. However, the continuum shape of the GCXE gives the bremsstrahlung temperature of $\sim$15\,keV, which is significantly larger than those of the GBXE and GRXE of $\sim$5~keV (table~\ref{tab:GDXE}). 
This apparent inconsistency in the temperatures would be due to different flux ratio  of the hard X-ray spectra associated to the \Ka\ line (cold gas component) relative to the high-temperature plasma associated with \Hea\ and \Lya\ line.
In fact, the flux ratio of \Ka/\Hea\ in the GCXE is $0.38\pm{0.02}$, which is significantly larger than those of the GBXE and GRXE of $0.20\pm{0.03}$ and  $0.27\pm{0.02}$, respectively.

\subsection{Sample of mCVs, non-mCVs, and ABs}  
\label{sec:AS}
We selected  the sample sources of intermediate polars (IPs), polars (Ps), symbiotic stars (SSs), non-mCVs (dwarf novae), and ABs from table~1 of \citet{Xu16}.
The IPs, Ps, and  SSs were combined into a single class of mCVs, because the number densities of Ps and SSs are  smaller than IPs \citep{Pa84}, and the X-ray spectra are similar to each other, compared with those of non-mCVs and ABs.  From the spectral features, we classify GK per to a mCV (IP) instead of a non-mCV, and omit BF~Ori from the list of non-mCV (see \citealt{Sh07, Ne08}).

For the ABs, we added other sources of Algol (OBSID$=$401093010),
EV~Lac (402032010),
HR~9024 (401032010),
HD130693 (405031010), and
$\beta$~Lyr (401036010, 401036020, 401036030).  
The sample names of mCVs, non-mCVs, and ABs in this paper are listed in table~\ref{tab:first}.

\input{List-AS.tex} 

\begin{figure}[hbt]
\figurenum{3}
\epsscale{1.0}
\plotone{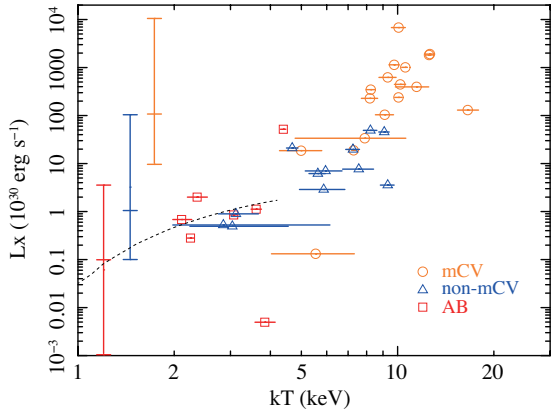}
\caption{
$kT$ and $L_{\rm X}$ plot of point sources.
The mCVs, non-mCVs, and ABs are shown with orange circle, blue triangle, and red square, respectively.
The dashed line shows a model curve of $kT$ and $L_{\rm X}$ calculated by APEC, which is used for ABs (see section~\ref{sec:modelB}).
The vertical bars show the groups we divided to make Model~B (see section~\ref{sec:modelB} and table~\ref{tab:ModelB}).}
\label{fig:kT-Lx}
\end{figure}

We make spectra of the samples, and fit with a CIE plasma (APEC) model. Free parameters are temperature $kT$, iron abundance $Z_{\rm Fe}$, and luminosity $L_{\rm X}$ in 5--10~keV.
We used distances of the point sources shown in table~1 of \cite{Xu16} (references therein) to convert fluxes into luminosities.
Figure~\ref{fig:kT-Lx} shows a positive correlation between $kT$ and $L_{\rm X}$.
The data of mCVs and non-mCVs distribute in the range of $L_{\rm X}\sim 10^{31-34}$~erg~s$^{-1}$ and $10^{29-32}$~erg~s$^{-1}$ corresponding to $kT\sim5$--$15$~keV and $3$--$10$~keV, respectively.
For ABs, except for two samples with large dispersion in $L_{\rm X}\lesssim 10^{28}$~erg~s$^{-1}$ and $\sim 10^{32}$~erg~s$^{-1}$, the luminosity $L_{\rm X}$ and temperature $kT$ are $10^{29-30.5}$~erg~s$^{-1}$ and 2--4~keV, respectively

\begin{figure}[phbt]
\figurenum{4}
\epsscale{1.0}
\plotone{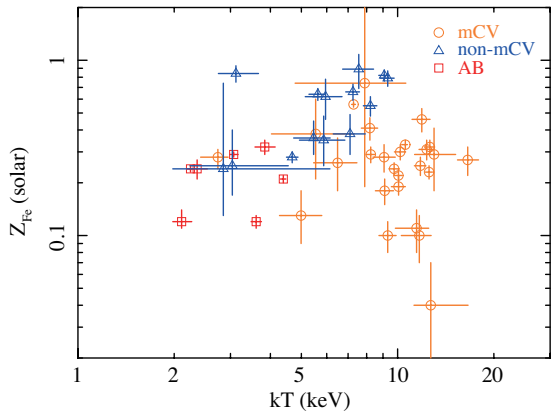}
\caption{
$kT$ and $Z_{\rm Fe}$ plot of point sources with the same symbols as figure~\ref{fig:kT-Lx}.
}
\label{fig:kT-ZFe}
\end{figure}
On the other hand, no clear correlation is seen between $kT$ and $Z_{\rm Fe}$  (figure~\ref{fig:kT-ZFe}).
The mean abundance $Z_{\rm Fe}$ and the standard deviation of mCVs, non-mCVs, and ABs are estimated to be $0.28\pm{0.16}$~solar, $0.58\pm{0.24}$~solar, and $0.22\pm{0.08}$~solar, respectively.  
The physical reason for this apparent difference of the mean abundances among mCVs, non-mCVs and ABs is not fully understood, but would be related to the different production mechanism of the hot plasmas; the hot plasmas are produced by the shock of a freefall velocity  on the white dwarf surface (mCVs),  Keplerian velocity near the white dwarf surface (non-mCVs), and coronal activity (ABs).

\subsection{Mean X-ray spectra of mCVs, non-mCVs, and ABs (Model A)} 
\label{sec:modelA}
\begin{figure*}[hbtp]
\figurenum{5}
\epsscale{0.33}
\plotone{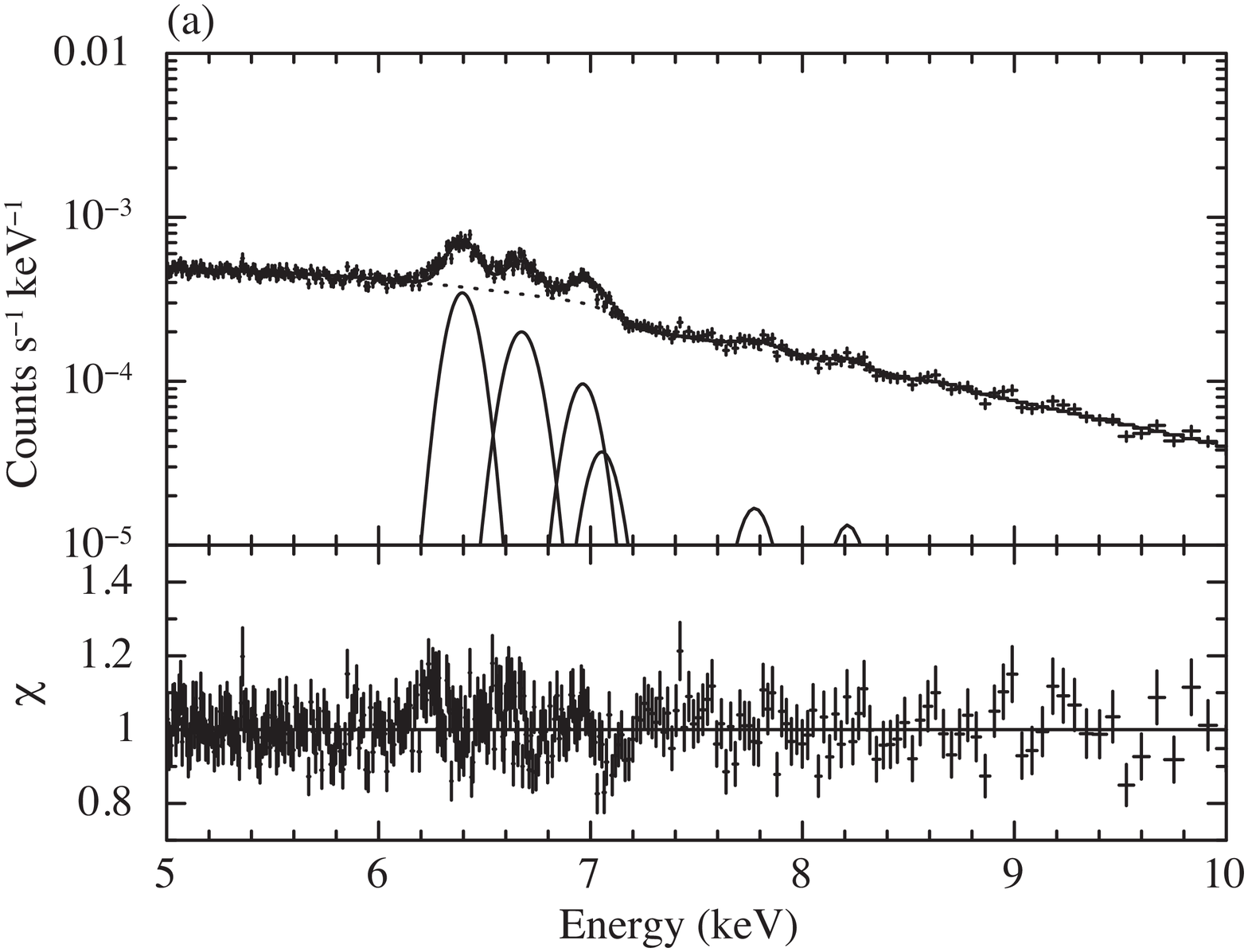}
\plotone{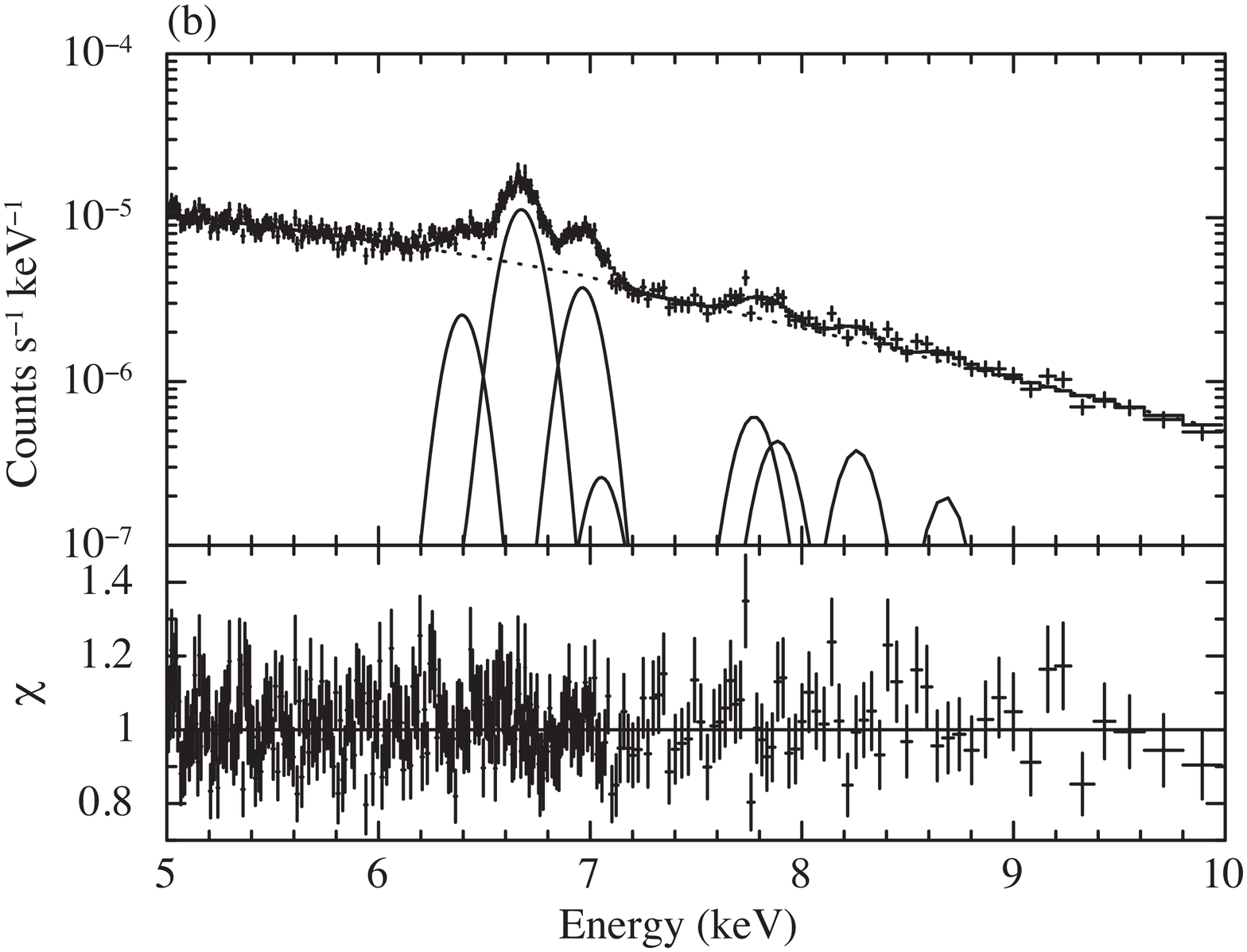}
\plotone{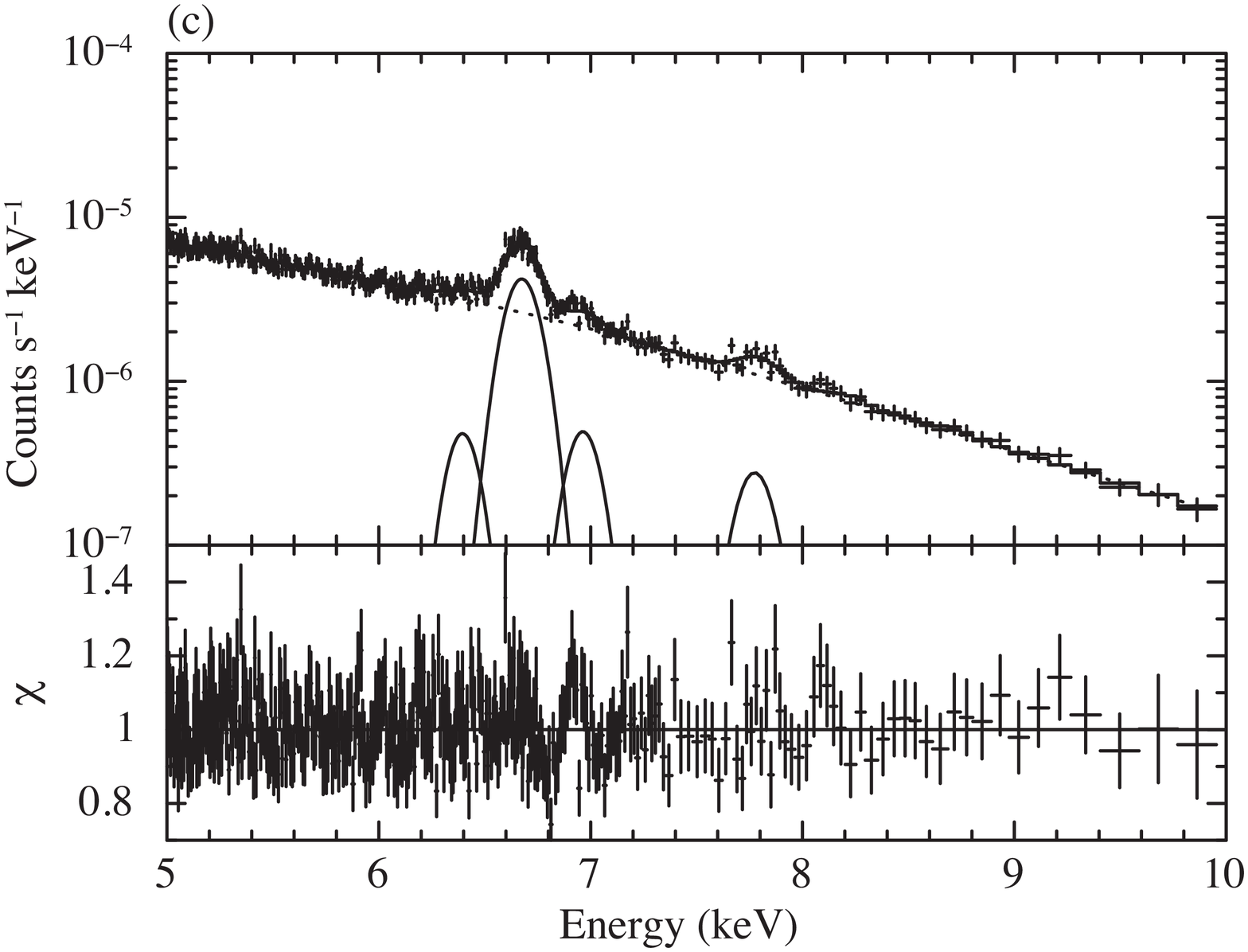}
\caption{
Averaged spectra of mCVs, non-mCVs, ABs for Model~A (see section~\ref{sec:modelA}), which are normalized as a point source located at 8~kpc.
}
\label{fig:ASspectra}
\end{figure*}

We co-added each source spectra in table~\ref{tab:first}, where each spectra are converted to those at the same distance of 8~kpc (hereafter, the mean spectra). The mean spectra of the mCVs, non-mCVs, and ABs are shown in figure~\ref{fig:ASspectra}. The CXB was subtracted from the spectra.
Then we fit the spectra by the same phenomenological model as section~\ref{sec:GDXEspectra}.
The best fit models are given in the solid line of figure~\ref{fig:ASspectra}, while the best-fit parameters are listed in table~\ref{tab:AS}. 
Parameters of SSs, IPs, and Ps, which are the subclasses of mCVs, are also shown in the table.
Hereafter, we call these mean spectra Model~A.

The Model A may not fully present the mean spectra of mCVs, non-mCVs, and ABs, because the  samples are limited in the luminosity ranges
of  $L_{\rm X}\sim10^{31-34}$\,erg\,s$^{-1}$,  $\sim 10^{29-32}$\,erg\,s$^{-1}$, and  $\sim 10^{29-30.5}$\,erg\,s$^{-1}$ (5--10~keV), for the mCVs, non-mCVs, and ABs, respectively (figure~\ref{fig:kT-Lx}). 
Contribution of samples with low luminosity is possibly underestimated due to the detection bias.
Since the sources in the luminosity range of $\lesssim10^{29-30}$\Lu~ may contribute non-negligible fractions in the mean spectra, we try to include the possible contribution of low-luminosity sources to the mean spectra of the mCVs, non-mCVs, and ABs in the next section (Model B).

\input{table-AS_all.tex} 

\subsection{Mean X-ray spectra of mCVs, non-mCVs, and ABs (Model B)} 
\label{sec:modelB}
Since the \Hea, and \Lya\ structure in the mCV, non-mCV, and AB spectra are well represented by a thermal plasma (e.g., mCVs: \citealt{Yu12}, non-mCVs: \citealt{By10}, ABs: \citealt{Pa12}), we fit the observed spectra of individual sources with a model of CIE plasma (APEC) plus \Ka, \Kb, and \NiKa\ lines given by Gaussians.  Using the best-fit model, we estimated \EWHea\ and \EWLya.

\begin{figure}[hbt]
\figurenum{6}
\epsscale{1.0}
\plotone{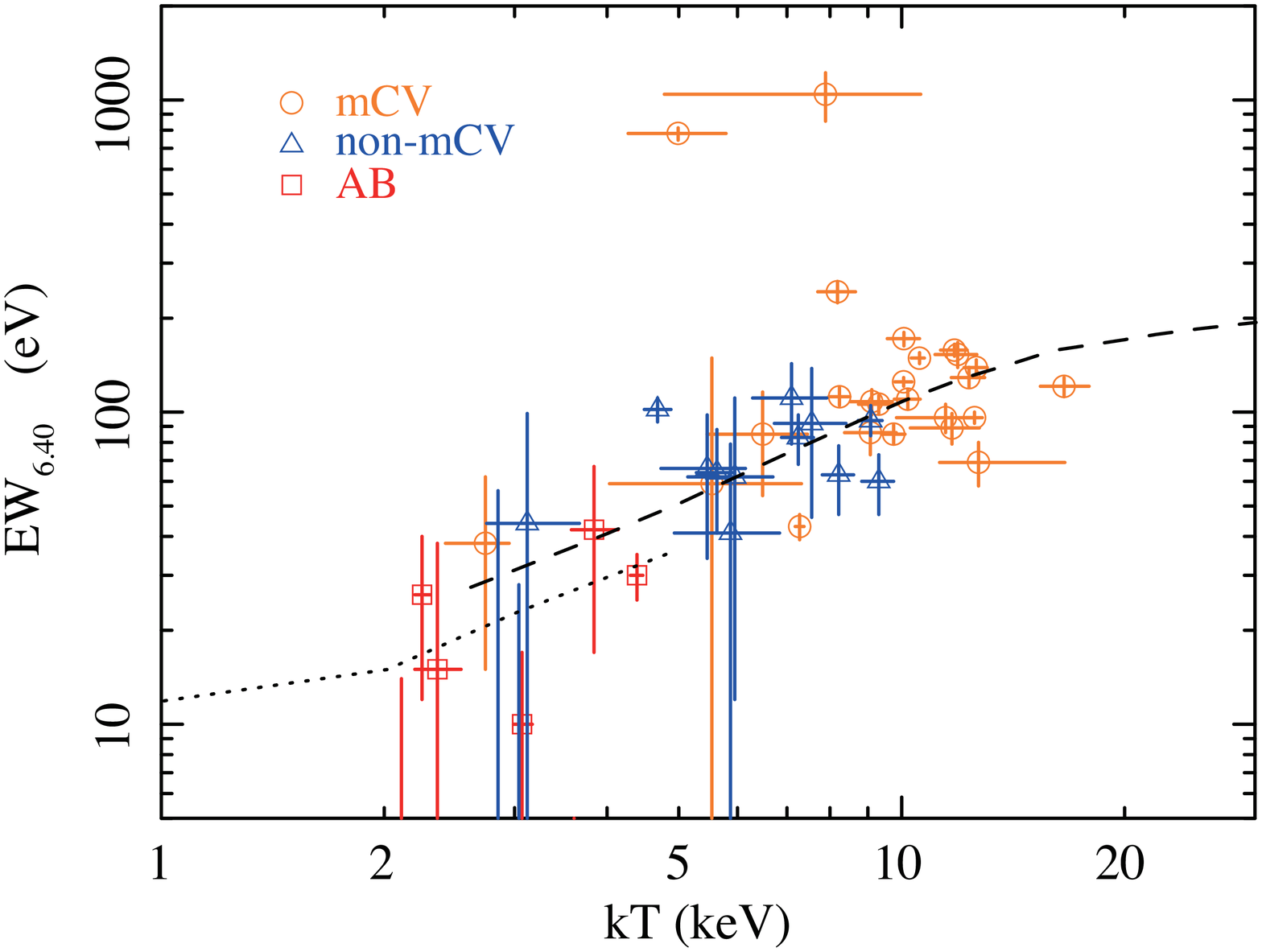}
\caption{
$kT$ and \EWKa\ plot of point sources with the same symbols as figure~\ref{fig:kT-Lx}.
The dashed and dotted curves show simulated results, where \EWKa\ is produced by irradiated X-rays from a plasma on the cold gas
with $N_{\rm H}=1\times10^{23}$~cm$^{-2}$ and $3\times 10^{22}$~cm$^{-2}$ for CVs (mCVs + non-mCVs) and ABs, respectively. 
}
\label{fig:kT-EWKa}
\end{figure}
Figure~\ref{fig:kT-EWKa} shows a correlation plot between $kT$ and \EWKa.
The \EWKa\ depends on the geometry and $N_{\rm H}$ of the cold gas near and around the hot plasma.  If we assume that the geometry and $N_{\rm H}$ of the cold gas around  the hot plasma  are the same in all the sources, the \EWKa\ becomes a simple function of $kT$; \EWKa\ is roughly proportional to the \Ka\ flux divided by the fluxes at above the Fe-K edge energy.  The dotted and dashed  lines in figure~\ref{fig:kT-EWKa} are the simulated results of the $kT$ vs \EWKa\ relation,  where \EWKa\ is produced by the irradiation of X-rays from a plasma on the cold gas
with temperature of $kT$ and gas density of $N_{\rm H}$. 
As is shown in figure~\ref{fig:kT-EWKa}, data of CVs (mCV+non-mCV) and ABs are well fitted with the model curve of $N_{\rm H}=1\times10^{23}$~cm$^{-2}$ (dashed line) and $3\times 10^{22}$~cm$^{-2}$ (dotted line), respectively. 

\begin{figure}[hbt]
\figurenum{7}
\epsscale{1.0}
\plotone{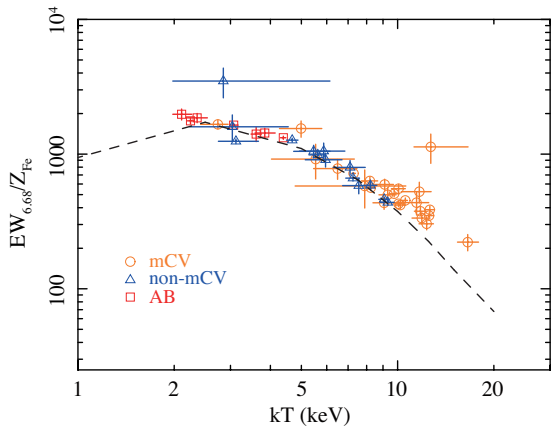}
\caption{
$kT$ and {\EWHea}/Z$_{\rm Fe}$ plot of point sources with the same symbols as figure~\ref{fig:kT-Lx}.
A simulated model that is derived from the APEC code is shown with the dashed curve.
}
\label{fig:kT-EWHea}
\end{figure}

The correlation plot of \EWHea\ divided by $Z_{\rm Fe}$ as a function of $kT$ is given in figure~\ref{fig:kT-EWHea}.
The dashed curve is a simulated result for a thermal plasma.  The results well reproduce the observed \EWHea/$Z_{\rm Fe}$, which supports our initial assumption that the spectra of mCV, non-mCV, and ABs are all well approximated  by a simple CIE plasma model at least in the energy band of 5--10 keV. 

Using the results in figures~\ref{fig:kT-Lx} and \ref{fig:kT-EWKa}, and adopting the mean $Z_{\rm Fe}$ of each category (section~\ref{sec:AS}), we can incorporate the XLF effect into the mean spectra  as a form of multi-$kT$, $Z_{\rm Fe}$, and \EWKa\ spectra.  However, for simplicity, we construct a two-representative (two $kT$ and \EWKa) plasma as a good approximation of multi-temperature and \EWKa\ structure.

\begin{figure}[hbt]
\figurenum{8}
\epsscale{1.0}
\plotone{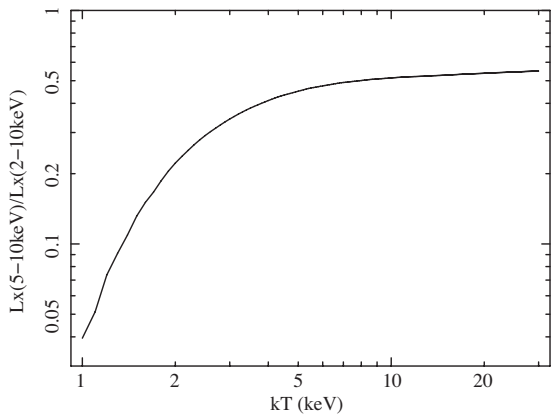}
\caption{
$kT$ and $L_{\rm X}$ ratio between 5--10~keV and 2--10~keV bands.
}
\label{fig:kT-LxRatio}
\end{figure}

\input{table-ModelB.tex}

For mCVs, we refer to the XLF provided by \citet{Wa14b}, where essentially all the source are in our range of $10^{31}$--$10^{34}$~erg~s$^{-1}$.
We are focusing on the 5--10~keV luminosity, which is different from the 2--10~keV band used by \cite{Wa14b}. 
We calculated conversion factor between the 2--10~keV and 5--10~keV luminosities using APEC model with various $kT$ (figure~\ref{fig:kT-LxRatio}), and modified the XLF for the 5--10~keV band (hear and after the modified XLF).
We, then, divided mCVs data into two groups of $10^{31-32}$~erg~s$^{-1}$ and $10^{32-34}$~erg~s$^{-1}$. 
The equivalent $kT$ ranges are 7 and 10~keV  (see figure~\ref{fig:kT-Lx}). 
We also calculated that the luminosity ratio between these luminosity bands are $0.98:1$ based on the modified XLF. 
Taking these representative $kT$s, we made a two-$kT$ and \EWKa\ spectrum of mCVs (Model B) as [$kT=7$~keV, \EWKa$=80$~eV] and [$kT=10$~keV, \EWKa$=110$~eV]  with the luminosity ratio of $0.98:1$ (table~\ref{tab:ModelB}). 

In the same manner, the Model~B spectrum for the non-mCVs is made with two components of the luminosity ranges of $10^{29-30}$~erg~s$^{-1}$ and $10^{30-32}$~erg~s$^{-1}$ with [$kT=3$~keV, \EWKa$=35$~eV]  and [$kT=8$~keV, \EWKa$=90$~eV], respectively. 
Using the modified XLF,
we calculated the luminosity ratio of $0.17:1$.

The XLF of ABs is highly uncertain, but our sample of ABs would be in the highest luminosity range of $10^{29-30.5}$~erg~s$^{-1}$, where the relevant temperature range is $\sim$2--4\,keV. 
We make two groups of $L_{\rm X}=10^{27-29}$~erg~s$^{-1}$ and $10^{29-30.5}$~erg~s$^{-1}$.
Since there is no sample in the lower luminosity group, we assume the $kT$--$L_{\rm X}$ relation using the CIE plasma model (APEC) (the dashed curve in figure~\ref{fig:kT-Lx}).
Then, the two groups have parameter sets of  [$kT=1$~keV, \EWKa$=10$~eV] and  [$kT=3$~keV, \EWKa$=25$~eV], respectively.
The luminosity ratio $1.6\times 10^{-2}:1$ is also calculated
 from the modified XLF in the same manner as mCVs and non-mCVs.
Although contribution of low-$kT$ ABs of $\lesssim$2~keV may not be negligible in the 2--10 keV flux (e.g. \citealt{Sa06, Wa14b}), the contribution to in the 5--10 keV band would be very small. This is because low-temperature plasma hardly emits high energy X-rays of $\gtrsim5$~keV. 
Using the parameters listed in table~\ref{tab:ModelB}, we constructed Model B for mCVs, non-mCVs, and ABs.

\subsection {Reconstruction of the GDXE spectra by assembly of point sources} 
\label{sec:reconstruction}
For the reconstruction of the GDXE by assembly of point sources (mCVs, non-mCVs, and ABs) , key parameters are EWs of iron (\Ka, \Hea, and \Lya). 
In the next subsections~\ref{sec:GBXE} and \ref{sec:GRXE-GCXE}, we simply reconstruct the observed GDXE spectra with those of the observation-based model (Model A).
However, for the reconstruction by Model B, the iron abundances in the ISMs between the solar neighbor and that near the GDXE should be taken into account,
because the observed iron EWs would be proportional to the iron abundances
in the relevant ISM. The iron abundance in the GDXE is nearly 1 solar in the X-ray observations \citep{Ko07b,No10,Uc13}.
Also the infrared observation gives nearly 1 solar iron abundance in the ISM near the Galactic center \citep{Cu07}. 
Therefore, for the reconstruction of the GDXE by Model B, we assume the iron abundances in the solar neighborhood and that in the GDXE regions are the same, being 1 solar, and hence no abundance correction of the iron EWs in both the GDXE and Model B is made.

\subsubsection {GBXE} \label{sec:GBXE}
\begin{figure*}[hbt]
\figurenum{9}
\epsscale{0.45}
\plotone{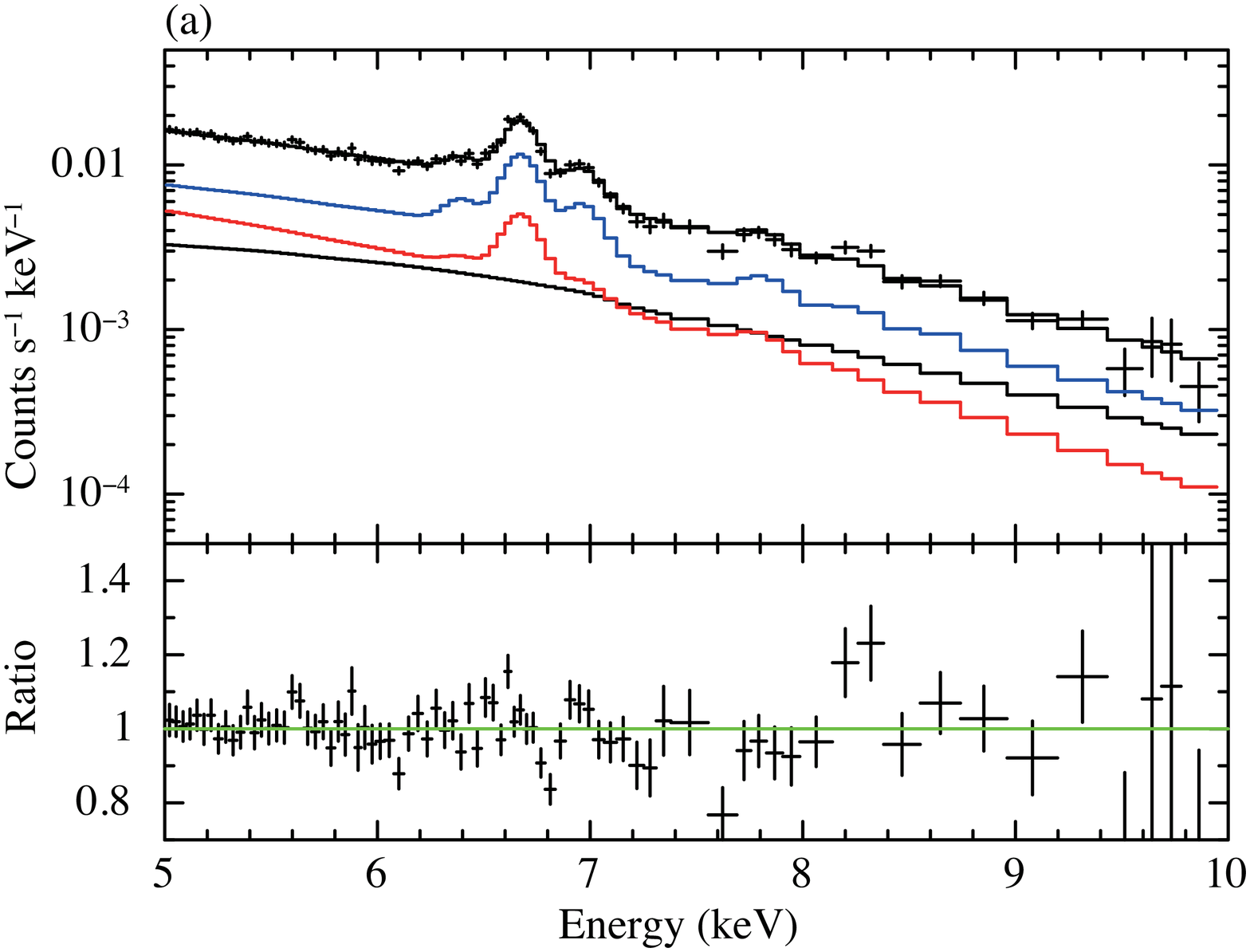}
\plotone{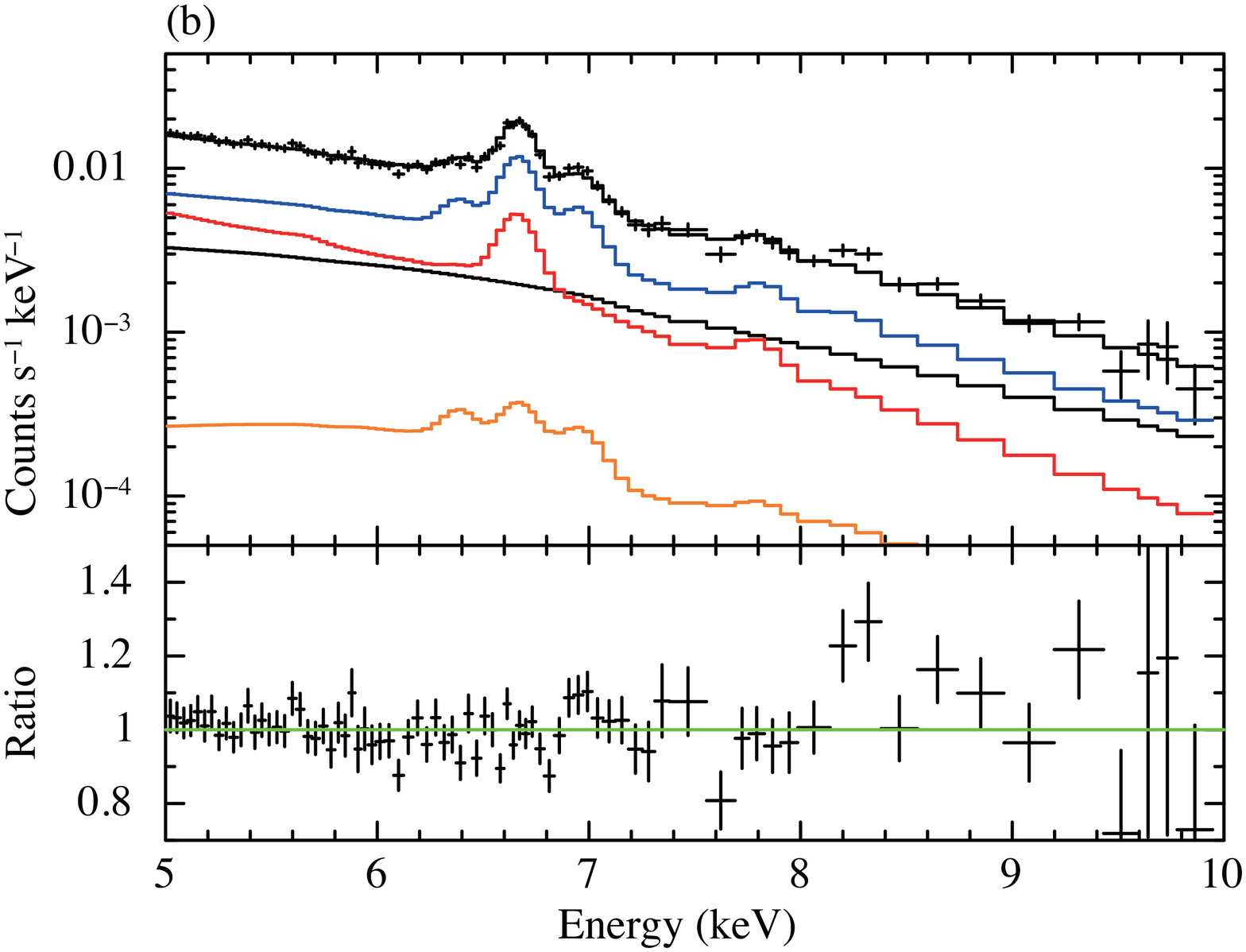}
\caption{
(a) GBXE spectrum fitted with the combination of the mCVs (orange), non-mCVs (blue), and ABs (red) of the Model A. In this case, mCVs rarely contributes the 5--10~keV spectrum. 
The black solid curve shows the CXB model.
(b) Same as the left panel, but fitted with the Model~B.
}
\label{fig:GBXEspectra}
\end{figure*}

\citet{Re09, Ho12} suggested that more than 80\% of the GBXE flux is due to integrated point sources of CVs (mCVs and non-mCVs) and ABs. 
We, therefore, fit the GBXE spectrum with a combination of the mean spectra of mCVs, non-mCVs and ABs (Model A and Model B) plus the fixed CXB model. 
The free parameters are $F_{\rm mCV}, F_{\rm non-mCV}$, and $F_{\rm AB}$ (\Fu~arcmin$^{-2}$), which are the surface brightness of mCVs, non-mCVs, and ABs, respectively.  Here, we define the parameters of fraction, $f_{\rm mCV}=F_{\rm mCV}/Sum$, $f_{\rm non-mCV}=F_{\rm non-mCV}/Sum$ and  $f_{\rm AB}=F_{\rm AB}/Sum$,  where $Sum$ is $F_{\rm mCV}$+$F_{\rm non-mCV}$+$F_{\rm AB}$.
The interstellar column density of \NH\ is fixed to $3\times10^{22}$~cm$^{-2}$.
The \NH~value, however, has no significant effect on  the best-fit parameters of the 5--10 keV band spectra.  The  fits  are  reasonably nice with $\chi^2$/d.o.f.$=160/95$ and $148/95$  for the Model~A and Model~B, respectively, although residuals at $\sim 8.3$~keV are seen. The best-fit results are shown in figure~\ref{fig:GBXEspectra}, and  the mixing ratio for Model~B is listed in table~\ref{tabl:fit-ModelB}.

The fitting result suggests that the major fraction of the GBXE is due to non-mCVs.
No essential difference between the Model A and Model B fit is found.
We, thus, use Model~B for the GRXE and GCXE spectra in the next subsection.

\subsubsection{GRXE and GCXE} \label{sec:GRXE-GCXE} 
\begin{figure*}[thbp]
\figurenum{10}
\epsscale{0.45}
\plotone{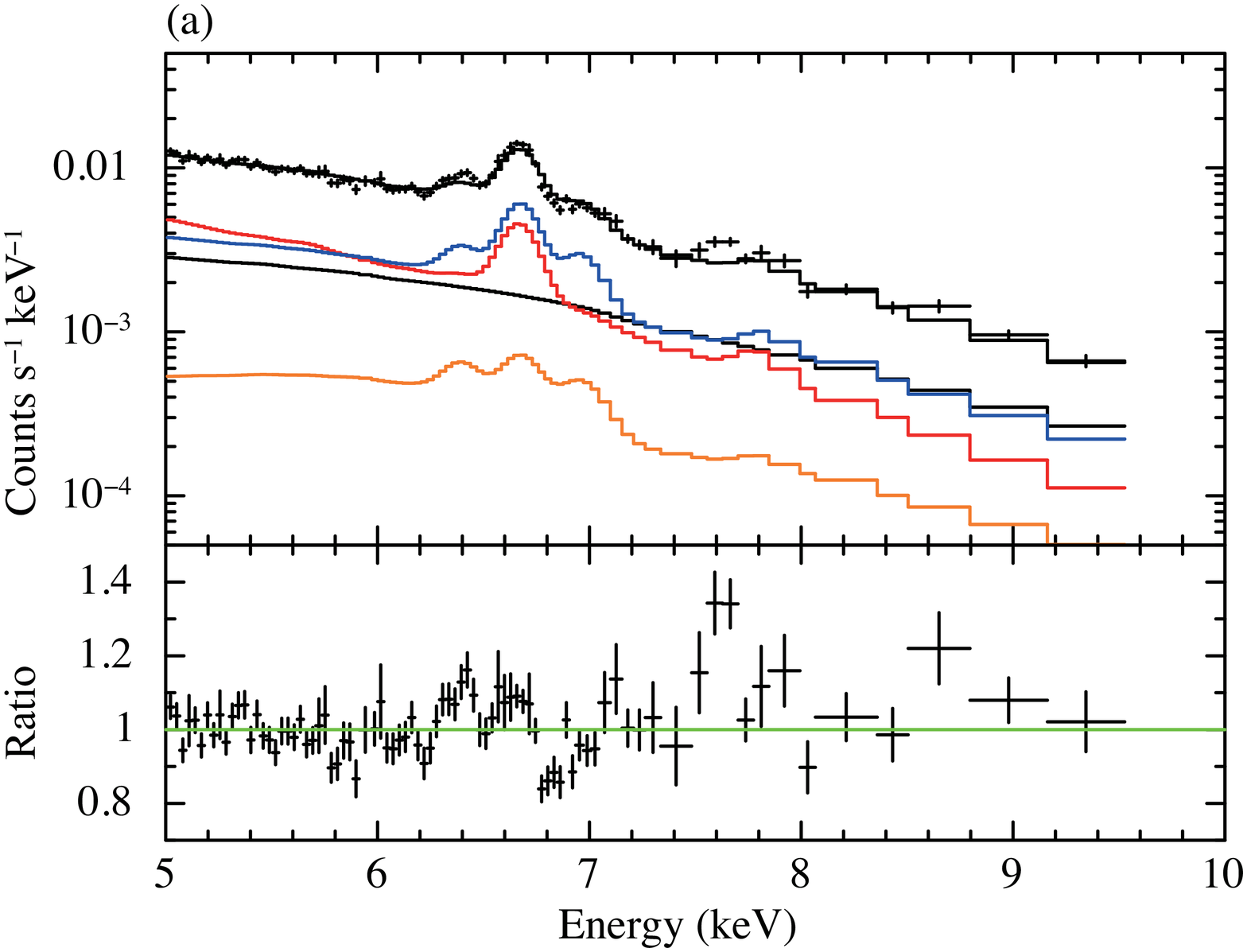}
\plotone{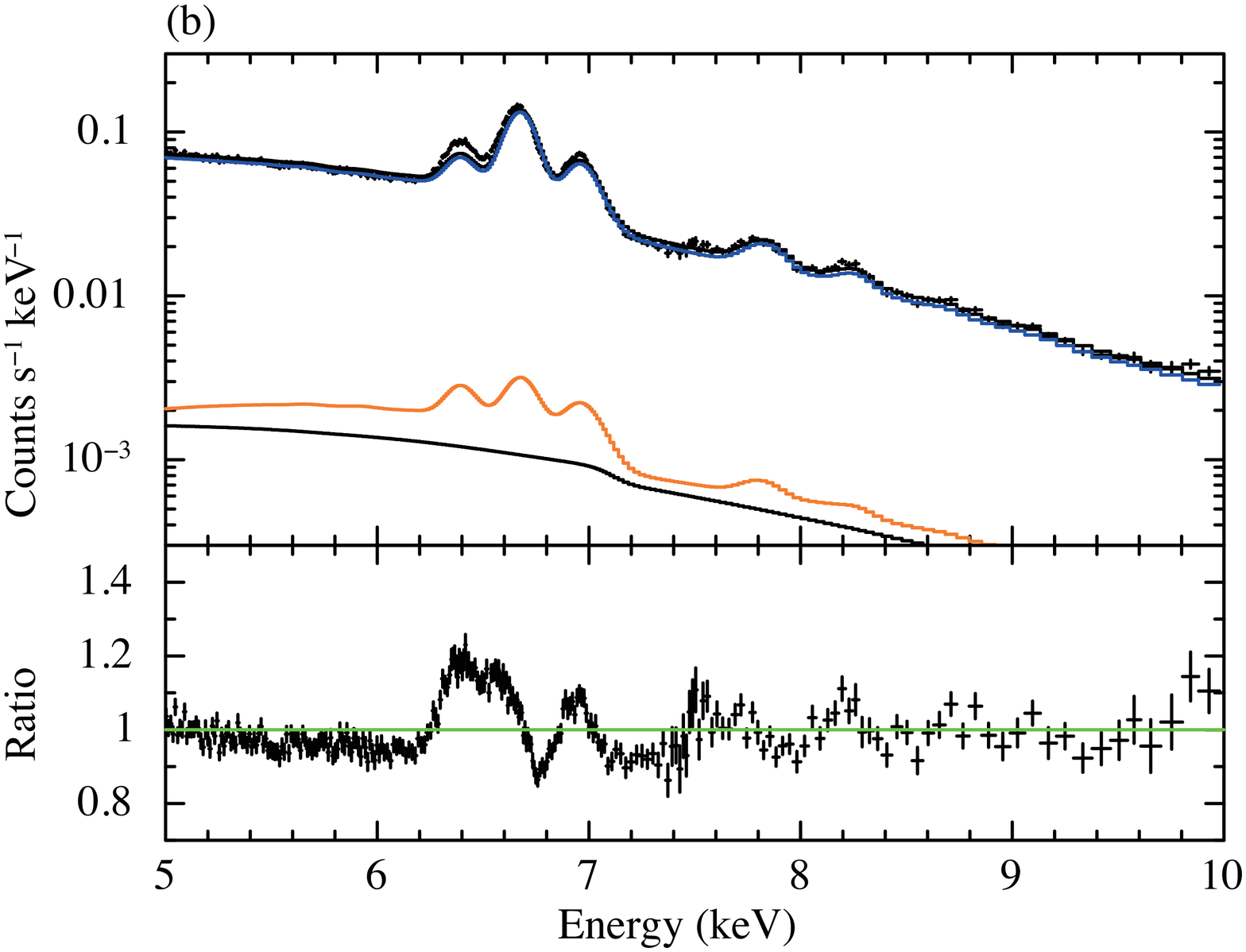}
\caption{
(a) GRXE spectrum fitted with the combination of the mCVs (orange), non-mCVs (blue), and ABs (red) of the Model~B.
 The black solid curve shows the CXB model. 
(b) Same as (a), but for the GCXE.
}
\label{fig:GCXE_GRXEspectra}
\end{figure*}

Unlike the GBXE, there are no observational facts to resolve the majority of the GRXE and GCXE flux into point sources.  
We, nevertheless, try to fit the GRXE and GCXE spectra by a combination of the mean spectra of mCVs, non-mCVs, and ABs (figure~\ref{fig:GCXE_GRXEspectra}).
The free parameters are the same as the GDXE but \NH\ are fixed to $6\times10^{22}$~cm$^{-2}$ and $3\times10^{22}$~cm$^{-2}$, for the GRXE and GCXE, respectively.  
The fitting rejects that the GRXE and GCXE spectra are combinations of only
mCVs, non-mCVs, and ABs, with the $\chi^2$/d.o.f. of $282/91$, and $2637/276$ for the GRXE and GCXE, respectively, where large residuals are found in the 6.2--7.2 keV band. 
The Model~B fitting also retains residuals at $\sim 7.6$~keV in the GRXE spectrum, and at $\sim 7.5$--7.9~keV and $\sim8.2$--8.3~keV in the GCXE spectrum, respectively.

\input{FluxRatio.tex} 

The main disagreement of the GRXE and GCXE spectra  from the  combined model of  mCVs, non-mCVs, and ABs lie in the 6.2--7.2 keV band. This band includes the \Ka, \Hea, and \Lya~lines, and hence these lines are key elements to separately diagnose the origins of the GCXE, GBXE, and GRXE (sections~\ref{sec:dis-GBXE}, \ref{sec:dis-GRXE}, and \ref{sec:dis-GCXE}). 

\section{Discussion} \label{sec:discussion} 
\subsection {Equivalent Widths of the iron K-shell lines from mCVs, non-mCVs, ABs, and GDXE} 
\label{sec:dis-EW}
We have determined the mean  \EWKa, \EWHea, and \EWLya, with respective values of $169\pm{5}$~eV, $118\pm{5}$~eV, and $60\pm{4}$~eV for mCVs, $82\pm{7}$~eV, $451\pm{10}$~eV, and $167\pm{9}$~eV for non-mCVs, and $28\pm{5}$~eV, $327\pm{8}$~eV, and $45\pm{6}$~eV for ABs, respectively (Model A, table~\ref{tab:AS}). These EWs are consistent with, but are more accurate with smaller errors than the previous reports.  We further detected the \NiHea, \Heb, and \Lyb~lines  in the mCVs spectrum, and \NiHea\ and  \Heb\ lines in  the non-mCVs and ABs spectra for the first time.  

We have also obtained high-quality spectra of the  GDXE, and accurately determined \EWKa, \EWHea, and \EWLya.  In addition, we detect many K-shell lines of iron and nickel such as \NiKa, \NiHea, \Heb, \Lyb, \Hec\ and \Lyc~lines from the GCXE spectra.
From the GBXE and GRXE spectra, newly detected lines are \NiHea, \Hec, and \Lyc. 

The \EWKa, \EWHea, and \EWLya~of  mCVs, non-mCVs, and ABs  have been reported by several authors, mainly with  {\it ASCA}, {\it Chandra}, {\it XMM-Newton} and {\it Suzaku} (\citealt{Ya16} and references therein). 
However, the mean values of the EWs have large errors, except for \citet{Xu16}, and have large variations from author to author and/or from instrument to instrument.  
These large systematic errors would be due to different analysis processes from author to author for the rather faint iron K-shell structures: different instrument, different criteria of the data selections, reductions, the NXB estimations, different analysis tools, and the various other conditions.

We have estimated the EWs using the same instrument (XIS) with unified  data reduction and analysis for all the GDXE, mCVs, non-mCVs, and ABs spectra.
Therefore, the systematic errors of EWs, in particular relative systematic errors of EWs among the GCXE, GRXE, GBXE, mCVs, non-mCVs, and ABs would be far smaller than those of the previous reports.  
This is essential for the reliable reconstruction of the GCXE, GRXE, and GBXE spectra by the combination of the mean spectra of mCVs, non-mCVs, and ABs.
  
\subsection{Galactic Bulge X-ray Emission (GBXE)}
\label{sec:dis-GBXE}
Using the deep {\it Chandra} observations ($\sim$1~Ms) at  $(l_*, b_*)=(0\degr.1, -1\degr.4)$, \citet{Re09, Ho12} made the plot of the integrated  point source flux (6.5--7.1 keV) as a function of the threshold  luminosity (2--10 keV). They concluded that more than $\sim$80\% flux of the central region is resolved into point sources (figure~3b of \citealt{Ho12}). However, a problem is that number fractions (observed XLF) of CVs and ABs are significantly different between these two authors.

The high-quality GBXE and point source (mCVs, non-mCVs, and ABs) spectra with accurate \EWKa, \EWHea, and \EWLya\ obtained in this paper enable us to adopt a different approach to the point source origin for the GBXE.  
The GBXE spectrum, particularly the \EWKa, \EWHea, and \EWLya and relative ratio, are well reproduced by the combined model of mCVs, non-mCVs, and ABs (figure~\ref{fig:GBXEspectra}). 
This is consistent with the point source origin proposed by \citet{Re09, Ho12}.  The major fraction is occupied by non-mCVs,
in contrast to \cite{Re09, Ho12}.

The scale heights (SHs) of the \Ka\ (\SHKa),  \Hea\ (\SHHea) and the \Lya\ (\SHLya) are $\sim150$~pc,  $\sim$300~pc, and $\sim$300~pc, respectively \citep{Ya16}.
These SHs are consistent with  those of the mCVs, non-mCVs, and ABs, which also supports that the origin of GBXE is assembly of point sources, mainly non-mCVs (for \Hea, \Lya, and the 5-10 keV band flux), partly mCVs (for \Ka) and ABs (for \Hea).
The residual at $\sim 8.3$~keV corresponds to \Hec\ and/or \Lyb,  which will be  discussed in section~\ref{sec:dis-GCXE}.


\subsection {Galactic Ridge X-ray Emission (GRXE)} 
\label{sec:dis-GRXE}
\citet{Eb05} resolved $\sim$10\% of the GRXE flux into point sources above the detection threshold of $\sim2\times10^{31}$~\Lu~(2--10~keV) in the deepest observation of the GRXE field at $(l_*, b_*)\sim(28\degr.5, -0\degr.0)$.
The resolved fraction of the same region by \citet {Re07a} is about 20\% above the detection threshold of $\sim10^{31}$~\Lu~(1--7~keV). These differences are hard to be absorbed by the difference of the detection threshold, and hence may be regarded as a systematic error in the estimation of point source fraction. 

We have obtained the high-quality GRXE spectrum, which includes the \Ka, \Hea, \Lya, \Kb, \NiHea, \Hec, and \Lyc\ lines.  Unlike the GBXE, the GRXE spectrum cannot be well fitted with any combination of mCVs, non-mCVs and ABs ($\chi^2$/d.o.f.$=282/91$).  Large excesses are found at the \Ka\ and \Hea\ lines (figure \ref{fig:GCXE_GRXEspectra}a).  
Taking the continuum flux into account, we estimated that the \Ka\ and \Hea\ fluxes of the GRXE are $\sim 2$ and $1.2$ times larger than that estimated from the best-fit combined model of mCVs, non-mCVs, and ABs. 

The scale height of \SHKa, \SHHea, \SHLya, and that of the 5--10 keV band flux  of the GRXE  are significantly smaller than those of the GBXE,  and are inconsistent or  marginal to those of mCVs, non-mCVs, and ABs \citep{Ya16}.  These are consistent with that the GRXE cannot be reproduced by any combination of  mCVs, non-mCVs nor ABs.  The largest residual is found at \Ka~line.  The \SHKa~of the GRXE of  $\sim$70~pc is similar to molecular clouds \citep{Ya16}.  Since the \Ka~line shows local enhancements on the Galactic ridge, they argued that a significant fraction of  \Ka~is due to local bombardment of the low-energy cosmic rays (LECRs) to the molecular clouds.  
As for the origin of LECRs, \citet{Ta99} proposed reconnections of the magnetic field which is magnified by the possible turbulent motion of gas in the Galactic ridge.  This process may also produce a high-temperature plasma, and hence would compensate  the deficiency of the \Ka\ and \Hea\ lines in the combined model (figure~\ref{fig:GCXE_GRXEspectra}a).
The origin of the residual at $\sim7.6$~keV is discussed in section \ref{sec:dis-GCXE}. 


\subsection {Galactic Center X-ray Emission (GCXE)}  
\label{sec:dis-GCXE}
In the deepest observation ($\sim$600 ksec exposure) near the Galactic center of the $17^\prime \times 17^\prime$ field around Sgr A$^*$ ($40\times40$~pc), \citet{Mu03} resolved 
$\sim$10\% of the total flux (2--8 keV band) into point sources above the threshold luminosity of $\sim10^{31}$~\Lu. In the region $2^\prime$--$4^\prime$ west from 
Sgr A$^*$ ($\sim$900 ksec), \citet{Re07b} resolved $\sim$40\% of the flux (4--8 keV band) into point sources above the same threshold luminosity ($\sim10^{31}$\Lu). 
This fraction corresponds to $\sim$25--30\% in the 2--8 keV band in the XLF of \citet{Re09}. Thus, these two results are inconsistent with each other, which may be either due  to spatial fluctuations or more likely due to large systematic errors in deriving the flux of very faint point sources, as are discussed in the previous sections. 

\citet{Uc11} reported that the longitude profile of the  \Hea\ line flux is at least two times larger  than that of the SMD with the assumption that all the GRXE and GBXE are due to point sources (stars). The SMD  is  determined by the infrared observations made from the COBE/DIRBE data in the LAMBDA archive\footnote{http://lambda.gsfc.nasa.gov}, IRAS and IRT \citep{La02, Mu06}. 
The similar \Hea\ excess in the GCXE region is confirmed by the direct infrared star-counting observation of the SIRIUS by \citet{Yasui15}.

The \EWKa, \EWHea\ and \EWLya\ of the GCXE are all larger than those of mCVs, non-mCVs, and ABs (tables~\ref{tab:AS} and \ref{tab:GDXE}). 
In fact, the fit of the GCXE spectrum  by a combination of the mCV, non-mCV and AB spectra is completely rejected with the large excesses in the \Ka, \Hea,  and  \Lya\ lines (figure \ref{fig:GCXE_GRXEspectra}b). 
The \Ka, \Hea, and \Lya\ fluxes of the GCXE (table~\ref{tab:GDXE}) are, respectively, $\sim$2.0, 1.2, and 1.3 times larger than those estimated from the best-fit combined model (Model~B).  
These excess ratios (relative flux) are larger than possible systematic relative errors, and hence the excesses of iron lines are robust results. 

The \Ka~line is due to cold gas, while the \Hea\ and \Lya\ lines are attributable to hot plasma. Thus, a significant contribution of additional components is required regardless of whether diffuse or other point sources.  
This component should emit stronger K-shell lines of iron than any other known categories, and simultaneously satisfy apparently opposite 
characteristics: excess of cold gas (\Ka) and  that of  hot plasma (\Hea\ and \Lya).
  
The scale height of \SHKa, \SHHea, \SHLya, and that of the 5--10~keV band flux  of the GCXE are all similar to $\sim30$--$35$~pc \citep{Ya16}.  These are far smaller than those of mCVs, non-mCVs and ABs, and are 
more like that of  the central molecular zone (CMZ) \citep{Ts99, Wi15}.  Therefore, the origin of the GCXE may be closely related to the CMZ. Near 
Sgr A$^*$ ($|l_{*}|\lesssim 0\degr.3$), the longitude profile of \Hea~in the east (positive $l_{*}$) shows a significant excess over the west, even if we 
exclude the bright supernova remnant Sgr A East. This excess would be due to larger populations of high-mass stars in the east than the west \citep{Pa04, Mu04a, Ko07d}. 
These high-mass stars may contribute to the GCXE by possible starburst activity and frequent supernova  in the CMZ.  Another  possibility would be
magnetic reconnection in the CMZ with strong magnetic field, similar to the GRXE \citep{Ta99}. 

Big out-bursts of Sgr A$^*$  in the past \citep{In09, Te10, Po10, Ca12} would make a very hot plasma and LECRs, which may make ionized irons higher than that in normal plasmas near the GC.  Then  the transitions of highly excited level ($n>$2) to the ground state ($n=1$) are more enhanced compared to the  CIE plasma.  Line-like residuals at $\sim$7.8--7.9 keV  and $\sim$8.2--8.3 keV, would be such enhanced iron lines. \citet{Na13} discussed a possible effect of past big flares of Sgr A$^*$ on a plasma spectrum at the south of the GC. The residual at $\sim7.6$~keV found in the GCXE and GRXE corresponds to \NiKa, because unlike \Ka\ and \Kb, \NiKa\ is not included in the Model B spectra.

\section {Summary}
The summary for the origin of the GDXE based on the equivalent widths and scale heights of iron K-shell lines of the GCXE, GRXE, GBXE, mCVs, non-mCVs, and ABs are as follows;
\begin{itemize}

\item Equivalent width of the iron K-shell lines (\EWKa, \EWHea, and \EWLya) and their intensity patterns are different between the GCXE, GRXE and GBXE.

\item The X-ray spectrum near the iron K-shell lines of the GBXE is well explained by
the non-mCVs dominant plasmas with  small contribution of mCVs and ABs.

\item The X-ray spectrum near the iron K-shell lines of the GRXE shows significant excess at 6.4 keV line (\EWKa) from any combination of mCVs, non-mCVs, and ABs. The excess of \EWKa\ is likely due to low energy cosmic protons.

\item The X-ray spectrum of the GCXE shows significant excesses of \Ka, \Hea, and \Lya\ from any combination of mCVs, non-mCVs, and ABs. Therefore, significant fraction of GCXE is not due to mCVs, non-mCVs, and ABs.  A possible origin would be diffuse emission produced by high activity near the GC, such as the past big flares of Sgr A$^*$.
\end{itemize}

\bigskip
The authors are grateful to all members of the {\it Suzaku} team  who provide us with the excellent spectral data.  
This work is supported in part by the Grant-in-Aid for Scientific Research of the Ministry of Education, Science, Sports, and Culture (No. 15H02090, MN; No. 24540232, SY; No. 24540229, KK).
KKN is supported by Research Fellow of Japan Society for the Promotion of Science for Young Scientists.



\listofchanges

\end{document}

%% file: table-GCXE-Uc.tex
\begin{table*}[hbtp] 
\caption{The best-fit parameters of the GDXE.}
\label{tab:GDXE}
\smallskip
\footnotesize
\begin{center}
 \begin{tabular}{lccccccc}
       \hline
       \hline   
&&\multicolumn{2}{c}{GBXE} &\multicolumn{2}{c}{GRXE}&\multicolumn{2}{c}{GCXE}\\
\hline
\multicolumn{8}{c}{Continuum}\\
\multicolumn{2}{l}{\NH~($10^{22}$~cm$^{-2}$)} &\multicolumn{2}{c}{3 (fix)} &\multicolumn{2}{c}{3 (fix)} &\multicolumn{2}{c}{6 (fix)}\\
\multicolumn{2}{l}{Fe K edge$^{a}$} &\multicolumn{2}{c}{0 (fix)} &\multicolumn{2}{c}{0 (fix)} &\multicolumn{2}{c}{$0.24\pm0.01$}\\
\multicolumn{2}{l}{$kT_{\rm e}$ (keV)}&\multicolumn{2}{c}{$5.1\pm{0.4}$} &\multicolumn{2}{c}{$5.0\pm{0.4}$}&\multicolumn{2}{c}{$14.9^{+0.5}_{-0.6}$}\\
\multicolumn{2}{l}{flux~$^{b}$}&\multicolumn{2}{c}{$8.6\times10^{-15}$}&\multicolumn{2}{c}{$7.3\times10^{-15}$}&\multicolumn{2}{c}{$1.1\times10^{-13}$} \\
\hline
			\multicolumn{8}{c}{Emission lines}\\
Line$^c$	&CE$^c$  	& Flux$^d$ &EW$^e$		&Flux$^d$ &EW$^e$		&Flux$^d$ &EW$^e$ \\
  	\hline
\KaTab      &6400  	& 0.14$\pm{0.02}$&84$\pm10$ & 0.16$\pm{0.01}$&118$\pm9$&3.54$\pm{0.04}$&175$\pm2$\\
\HeaTab	&6680	&0.70$\pm{0.02}$&463$\pm13$ &0.60$\pm{0.02}$&487$\pm13$&9.40$\pm{0.05}$&500$\pm3$\\
\LyaTab    &6966	& 0.24$\pm{0.02}$&173$\pm13$&0.10$\pm{0.01}$&96$\pm11$&3.45$\pm{0.04}$&198$\pm2$\\
\KbTab	&7059       &	 0.01$^f$	     &14	 & 0.02$^f$           &19  &    0.44$^f$	         &26\\
\NiKaTab   &7490	&---	     	     &---		   &---		   &--- &0.37$\pm{0.07}$&24$\pm5$    \\
\NiHeaTab  &7771 	  &$0.05\pm{0.03}$	     &53$\pm31$	   &0.14$\pm{0.03}$&173$\pm36$ &1.10$\pm{0.06}$&78$\pm4$ \\
\HebTab 	&7881 	&$<0.06$	     & $<61$	   &$<0.06$& $<76$ &0.61$\pm{0.06}$&45$\pm4$  \\
\LybTab 	&8251 	&$<0.13$& $<101$ &$<0.09$	  &$<97$&0.74$\pm{0.12}$&54$\pm9$  \\
\HecTab	&8295 	&$0.11\pm0.08$	     & 130$^{+92}_{-98}$	   &$0.06\pm0.03$	  &$86\pm44$&0.38$\pm{0.13}$&30$\pm10$  \\
\LycTab 	&8700 	&0.09$\pm{0.03}$&120$\pm41$ &0.08$\pm{0.03}$&147$\pm49$ &0.34$\pm{0.05}$&30$\pm4$  \\
\hline
$\chi ^2$/d.o.f.  &    & \multicolumn{2}{c}{117/84}     & \multicolumn{2}{c}{107/72} & \multicolumn{2}{c}{331/265} \\
\hline
\end{tabular}
\end{center}
\tablenotetext{}{Errors are one standard deviation (1 $\sigma$).}
\tablenotetext{a}{Absorption depth at 7.11 keV.}
\tablenotetext{b}{Absorption-corrected flux in the unit erg~s$^{-1}$~cm$^{-2}$~arcmin$^{-2}$. The CXB is not included.}
\tablenotetext{c}{AtomDB 3.0.2 (http://www.atomdb.org/) and \cite{Wa05}. Unit is eV.}
\tablenotetext{d}{Unit is $10^{-7}$~photon~cm$^{-2}$~s$^{-1}$~arcmin$^{-2}$.}
\tablenotetext{e}{Unit is eV.}
\tablenotetext{f}{Fixed to 0.125$\times$\KaTab. }
\end{table*}

%% file: List-AS.tex
\begin{table*}
\caption{List of active stars used in this work.}
\smallskip
\footnotesize
\begin{center}
\begin{tabular}{l}
\hline
\hline
Magnetic CV (mCV) \\
\hline      
\ \ Symbiotic Stars(SS): CH~Cyg, RS~Oph, RT~Cru, SS73-17, T~CrB,  V407~Cyg \\
\ \ Intemediate polar(IP): AO~Psc,	BG~Cmi, EX~Hya, FO~Aqr, GK~Per, IGR~J17195$-$4100, IGR~J17303$-$0601, MU~Cam,\\
\ \ \ \ \ \ NY~Lup, PQ~Gem, TV~Col, TX~Col, V1223~Sgr, 1RXS~J213344.1$+$51072, V2400~Oph, V709~Cas, XY~Ari \\
\ \ Polar(P): AM~Her, V1432~Aql, SWIFT~J2319.4$+$2619\\
\hline
Non-magnetic CV (non-mCV) \\
\hline
\ \ BV~Cen, BZ~UMa, EK~TrA, FL~Psc, FS~Aur, KT~Per, SS~Aur, SS~Cyg, U Gem, V1159~Ori, \\
\ \ V893~Sco, VW~Hyi, VY~Aqr, Z~Cam\\
\hline
Active Binary (AB) \\
\hline
\ \ GT~Mus, 
Algol, II~Peg, $\sigma$~Gem, UX~Ari, EV~Lac, HR~9024, $\beta$~Lyr, HD~130693 \\
\hline
\end{tabular} \label{tab:first}
\end{center}
\tablenotetext{}{Observation sequence numbers of the sources are referred to \cite{Xu16} except for Algol, EV~Lac, HR~9024, HD130693, and $\beta$~Lyr (see text).}
\end{table*}

%% file: table-AS_all.tex
\begin{table*} 
\caption{Best-fit parameters of mCVs (SSs+IPs+Ps), SSs, IPs, Ps, non-mCVs, and ABs for Model A.}
\label{tab:AS}
\smallskip
\tiny
\begin{center}	
 \begin{tabular}{lccccccccccccc}
   \hline
 \hline   
&&\multicolumn{2}{c}{mCV} &\multicolumn{2}{c}{SS} &\multicolumn{2}{c}{IP} &\multicolumn{2}{c}{P} &\multicolumn{2}{c}{non-mCV}&\multicolumn{2}{c}{AB}\\
\hline
\multicolumn{14}{c}{Continuum}\\
\multicolumn{2}{l}{\NH~($10^{22}$~cm$^{-2}$)}&\multicolumn{2}{c}{14.7$\pm{1.5}$}&\multicolumn{2}{c}{18.0$\pm{2.3}$}&\multicolumn{2}{c}{9.7$\pm{0.4}$}&\multicolumn{2}{c}{12.7$\pm{3.4}$}&\multicolumn{2}{c}{0 (fix)}&\multicolumn{2}{c}{0 (fix)}\\
\multicolumn{2}{l}{Fe K edge$^{a}$} &\multicolumn{2}{c}{$0.05\pm0.01$}&\multicolumn{2}{c}{$<0.07$}&\multicolumn{2}{c}{$0.02\pm0.01$}&\multicolumn{2}{c}{$0.10\pm0.06$} &\multicolumn{2}{c}{$0.08\pm0.04$} &\multicolumn{2}{c}{$0.02(<0.05)$}\\
\multicolumn{2}{l}{$kT_{\rm e}$ (keV)}&\multicolumn{2}{c}{$23.3^{+5.1}_{-3.7}$}&\multicolumn{2}{c}{$24.5^{+5.3}_{-5.8}$}&\multicolumn{2}{c}{$18.7\pm0.6$}&\multicolumn{2}{c}{$25.7^{+20.7}_{-8.0}$}&\multicolumn{2}{c}{10.7$\pm{1.7}$}&\multicolumn{2}{c}{$4.25\pm{0.18}$}\\
\multicolumn{2}{l}{flux~(erg~s$^{-1}$~cm$^{-2}$)$^{b}$}&\multicolumn{2}{c}{$1.1\times10^{-13}$}&\multicolumn{2}{c}{$2.5\times10^{-13}$}&\multicolumn{2}{c}{$6.3\times10^{-14}$}&\multicolumn{2}{c}{$8.0\times10^{-15}$}&\multicolumn{2}{c}{$1.6\times10^{-15}$}&\multicolumn{2}{c}{$7.8\times10^{-16}$}\\
	\hline
			\multicolumn{14}{c}{Emission lines}\\
Line$^c$	&CE$^c$  & Flux$^d$ & EW$^e$ & Flux$^d$ & EW$^e$ & Flux$^d$ & EW$^e$ & Flux$^d$ & EW$^e$ & Flux$^d$ & EW$^e$ & Flux$^d$ & EW$^e$\\
				&				& $10^{-7}$ &				& $10^{-6}$ &				& $10^{-7}$ &				& $10^{-8}$ &				& $10^{-8}$ &				& $10^{-9}$ &			 \\
  	\hline
Fe\tiny{I}-K$\alpha$		&6400  	&4.88$\pm{0.13}$&169$\pm5$ 	&1.36$\pm{0.05}$&194$\pm7$ 	&1.90$\pm{0.04}$&124$\pm5$ 	&2.44$\pm{0.22}$&116$\pm10$ 	&0.26$\pm{0.02}$ &82$\pm7$		&0.49$\pm{0.09} $&28$\pm5$\\
Fe\tiny{XXV}-He$\alpha$	&6680	&3.19$\pm{0.12}$&118$\pm5$ 	&0.82$\pm{0.05}$&118$\pm7$ 	&1.63$\pm{0.04}$&114$\pm5$ 	&2.30$\pm{0.22}$&117$\pm11$ 	&1.29$\pm{0.03}$ &451$\pm10$ 	&4.97$\pm{0.12}$&327$\pm8$ \\
Fe\tiny{XXVI}-Ly$\alpha$    &6966	&1.52$\pm{0.11}$&60$\pm4$  	&0.37$\pm{0.04}$&60$\pm7$  	&0.89$\pm{0.03}$&67$\pm4$  	&0.70$\pm{0.20}$&38$\pm11$  	&0.44$\pm{0.02}$ &167$\pm9$ 	&0.60$\pm{0.09}$&45$\pm6$ \\
Fe\tiny{I}-K$\beta$		&7059   &0.61$^f$     	&25 					&0.17$^f$     	&25 					&0.24$^f$     	&18 					&0.30$^f$     	&17 					&0.07$^f$ &12 							&0.06$^f$  &5\\
Ni\tiny{I}-K$\alpha$  &7490	&-&-&-&-&-&-&-&-&-&-	&-&-\\
Ni\tiny{XXVII}-He$\alpha$ &7771 	&0.48$\pm{0.17}$ &23$\pm8$ 	&0.17$^{+0.05}_{-0.03}$ &34$^{+10}_{-6}$ 	&$<0.08$ &$<8$ 	&$<0.27$ &$<33$ 	&0.12$\pm{0.03}$ &56$\pm14$	&0.53$\pm{0.13}$&55$\pm13$\\
Fe\tiny{XXV}-He$\beta$ 	&7881 	&$<0.32$ &$<16$ 						&$<0.06$ &$<12$ 						&$0.18\pm0.05$ &$17\pm5$ 		&$0.50^{+0.23}_{-0.29}$ &33$^{+15}_{-19}$	&0.09$\pm{0.03}$ &43$\pm14$	&0.15$\pm{0.13}$ &17$\pm15$\\
Fe\tiny{XXVI}-Ly$\beta$ 	&8251 	&0.43$^{+0.13}_{-0.33}$ &22$^{+7}_{-17}$ 	&0.12$^{+0.05}_{-0.09}$ &25$^{+10}_{-19}$ 	&$<0.07$ &$<6$ 	&$<0.55$ &$<35$ 	& $<0.07$  & $<28$  					&0.16$\pm{0.10}$ &20$\pm14$\\
Fe\tiny{XXV}-He$\gamma$	&8295 	& $<0.35$ & $<18$ 					& $<0.08$ & $<17$ 					& $0.20^{+0.04}_{-0.07}$ & $20^{+4}_{-7}$ 					& $<0.58$ & $<40$ 					&0.09$^{+0.03}_{-0.07}$& 50$^{+17}_{-39}$	&$<0.08$&$<13$\\
Fe\tiny{XXVI}-Ly$\gamma$ 	&8700 	& $0.18\pm0.13$ & $10\pm7$ 	& $<0.10$ & $<24$ 	& $0.05\pm0.04$ & $5\pm4$ 	& $<0.43$ & $<33$ 	&0.06$\pm{0.03}$&34$\pm17$	&$<0.17$&$<24$\\
\hline
$\chi ^2$/d.o.f.  &   & \multicolumn{2}{c}{409/314}    & \multicolumn{2}{c}{363/302}    & \multicolumn{2}{c}{735/610}    & \multicolumn{2}{c}{109/130}    & \multicolumn{2}{c}{297/274}     & \multicolumn{2}{c}{361/363} \\
\hline
\end{tabular}
\end{center}
\tablenotetext{}{Errors are one standard deviation (1 $\sigma$).}
\tablenotetext{a}{Absorption depth at 7.11 keV.}
\tablenotetext{b}{Flux at 8~kpc in the 5--10~keV band per a source.}
\tablenotetext{c}{AtomDB 3.0.2 (http://www.atomdb.org/) and \cite{Wa05}. Unit is eV.}
\tablenotetext{d}{Unit is photon~cm$^{-2}$~s$^{-1}$.}
\tablenotetext{e}{Unit is eV.}
\tablenotetext{f}{Fixed to 0.125$\times$\KaTab. }
\end{table*}

%% file: table-ModelB.tex
\begin{table}[thp] 
\caption{Parameters of mCVs, non-mCVs, and ABs of Model~B.}
\label{tab:ModelB}
\smallskip
\footnotesize
\begin{center}	
\begin{tabular}{lcccccc}
\hline \hline
					&			&	$L_{\rm X}$			&	$kT$	& $EW_{6.40}$	& $Z_{\rm Fe}$	& $L_{\rm X}$ ratio$^a$	\\
					&			&	(erg~s$^{-1}$)		&	(keV)	& (eV)					& (solar)				& (low/high)	\\
\hline
mCVs			&	high	&	$10^{32-34}$		&	10		&	110					& 0.28					& ---	\\
					&	low	&	$10^{31-32}$		&	7			&	80					& 0.28					& $0.98$	\\
non-mCVs	&	high	&	$10^{30-32}$		&	8			&	90					& 0.58					& ---	\\
					&	low	&	$10^{29-30}$		&	3			&	35					& 0.58					& $0.17$	\\
ABs				&	high	&	$10^{29-30.5}$	&	3			&	25					& 0.22					& ---	\\
					&	low	&	$10^{27-29}$		&	1			&	10					& 0.22					& $1.6\times10^{-2}$	\\
\hline
\end{tabular}
\end{center}
\tablenotetext{a}{Luminosity ratio in the 5--10~keV band (see text).}
\end{table}

%% file: FluxRatio.tex
\begin{table} 
\caption{Best-fit flux ratio of mCVs, non-mCVs and ABs for Model B.}
\label{tabl:fit-ModelB}
\smallskip
\footnotesize
\begin{center}
 \begin{tabular}{lccc}
 \hline
 \hline
									&	GBXE	& GRXE	& GCXE	\\
\hline
$f_{\rm mCV}$			& $ 0.03 (<0.09)$	& $0.10\pm0.05$		& $0.04\pm0.01$ \\
$f_{\rm non-mCV}$	& $0.67\pm0.06$					& $0.51\pm0.06$		&	$0.96\pm0.01$ \\
$f_{\rm AB}$				& $0.30\pm0.03$					& $0.39\pm0.02$		&	$0.00 (<0.01)$ \\
\hline
$\chi^2$/d.o.f.	&	$148/95~(1.56)$	& $282/91~(3.10)$	& $2637/276~(9.55)$ \\
\hline
\end{tabular}
\end{center}
\end{table}